\documentclass[
	journal=jacsat,
	manuscript=article,
	]{achemso}

\newcommand{\figurewidth}{240pt}
\usepackage{siunitx}					
\DeclareSIUnit \rydberg {Ry}
\DeclareSIUnit \G {G_0}
\DeclareSIUnit \angstrom {\textsc{\AA}}

\usepackage{chemformula} 
\usepackage[T1]{fontenc} 

\usepackage{chemmacros}				
\newcommand{\bpy}{\iupac{4,4'-bi|pyri|dine}}
\newcommand{\silane}{\iupac{sil|ane|di|thio|methyl}}
\newcommand{\Silane}{\iupac{Sil|ane|di|thio|methyl}}

\usepackage[nolist,nohyperlinks]{acronym}
\begin{acronym}[ACRONYM]
	\acro{1DGH}[1DGH]{1D conductance histogram}
	\acro{cpd}[CPD]{change point detection}
	\acro{smbj}[SMBJ]{single-molecule break junction}
	\acro{pbe}[PBE]{Perdew-Burke-Ernzerhof}
	\acro{dz}[DZ]{double-$\zeta$}
	\acro{dzp}[DZP]{double-$\zeta$ polarized}
	\acro{gga}[GGA]{generalized gradient approximation}
	\acro{scf}[SCF]{self-consistent field}
	\acro{rt}[RT]{room temperature}
	\acro{bp}[BP]{break point}
	\acro{sm}[SM]{single-molecule}
	\acro{stm}[STM]{scanning tunneling microscope}
	\acro{si4}[Si4]{\silane}
	\acro{Si4}[Si4]{\Silane}
	\acro{bpy}[BPY]{\bpy}
	\acro{plh}[PLH]{plateau length histogram}
	\acro{2DH}[2DH]{2D histogram}
	\acro{ml}[ML]{machine learning}
	\acro{rss}[RSS]{residual sum of squares}
	\acro{dft}[DFT]{density functional theory}
	\acro{si}[SI]{Supporting Information}
	\acro{bj}[BJ]{break junction}
\end{acronym}

\usepackage[inline,shortlabels]{enumitem}

\newcommand{\UCPH}{Department of Chemistry and Nano-Science Center, University of Copenhagen, Universitetsparken 5, DK-2100, Copenhagen \O, Denmark}

\author{Joseph M. Hamill}
\affiliation{\UCPH}
\email{jmh@chem.ku.dk}

\author{William Bro-J\o{}rgensen}
\affiliation{\UCPH}

\author{Zolt{\'{a}}n Balogh}
\affiliation{Department of Physics, Institute of Physics, Budapest University of Technology and Economics, M{\H{u}}egyetem rkp. 3., H-1111 Budapest, Hungary}
\alsoaffiliation{HUN-REN-BME Condensed Matter Research Group, M{\H{u}}egyetem rkp. 3., H-1111 Budapest, Hungary}

\author{Haixing Li}
\affiliation{Department of Applied Physics and Applied Math, Columbia University, New York, New York 10027, United States}
\alsoaffiliation{Department of Physics, City University of Hong Kong, Kowloon 999077, Hong Kong SAR, China}

\author{Susanne Leitherer}
\affiliation{\UCPH}

\author{David H. Solomon}
\affiliation{Carroll School of Management, Boston College, 140 Commonwealth Avenue, Chestnut Hill, Massachusetts 02467}

\author{Andr{\'{a}}s Halbritter}
\affiliation{Department of Physics, Institute of Physics, Budapest University of Technology and Economics, M{\H{u}}egyetem rkp. 3., H-1111 Budapest, Hungary}
\alsoaffiliation{HUN-REN-BME Condensed Matter Research Group, M{\H{u}}egyetem rkp. 3., H-1111 Budapest, Hungary}

\author{Gemma C. Solomon}
\affiliation{\UCPH}
\alsoaffiliation{NNF Quantum Computing Programme, Niels Bohr Institute, University of Copenhagen, Belgdamsvej 17, DK-2100, Copenhagen \O, Denmark}

\title[]
{Improving single-molecule conductance measurements with change point detection from the econometrics toolbox}

\begin{document}

\begin{tocentry}
\includegraphics[width=3.25in,keepaspectratio,]{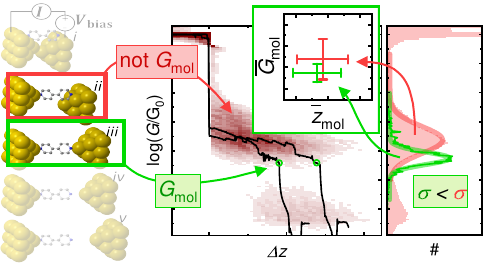}
\end{tocentry}

\begin{abstract}
Structural breaks occur in timeseries data across a broad range of fields, from economics to nanosciences.
For measurements of \aclp{smbj}, structural breaks in conductance versus displacement data occur when the molecular junction ruptures.
This moment is significant because the molecule is likely in its most extended geometry, and therefore resembles most closely the geometry used in theoretical predictions.
Conventional \acl{smbj} data analysis, on the other hand, typically uses the entire molecular plateau to estimate the single-molecule conductance, which skews the estimate when the plateau is sloped.
Borrowing from econometrics, where the study of structural breaks is well established, we present \ac{cpd} as a tool to search for junction rupture in \acl{smbj} data, and improve estimates in single-molecule conductance.
We demonstrate that using \ac{cpd} instead of the conventional \acl{1DGH} to determine the mean molecular conductance yields a standard deviation in the estimate of typically half that of the conventional approach, greatly improving accuracy.
We apply \ac{cpd} to three separate data sets, two on \acl{bpy} and one on a \acl{si4}, two at \acl{rt} and one at \qty{4}{\K}, two in one lab, one in another, to show the wide applicability of even the simplest of \ac{cpd} algorithms: the Chow test.
This versatility and better accuracy will propagate into more accurate theoretical simulations.
These improved metrics, in turn, will further improve any downstream analyses, including all emerging machine learning approaches.

\end{abstract}

\section{Introduction}

Measuring, understanding, and predicting \acl{sm} conductance is essential for a broad family of nanoscience fields.
\Acl{sm} sensing via nanopores use expected values of conductance to detect target molecular species for nanobiosensing applications.\cite{Veselinovic2019, Loh2018, Storm2003}
With \acl{sm} conductance measurements, \textit{in situ} detection and control of chemical reactions are also possible, including enzymatic processes.\cite{Yang2023, Mejia2019, Zang2019a, Ciampi2018, Huang2017, Aragones2016a, Besteman2003}
Two-state and three-state molecular switches correlate measured \acl{sm} conductance with geometric and electronic changes in target molecules due to electric fields and electrochemical potentials.\cite{Liljeroth2007, ODriscoll2017}
Furthermore, \acl{sm} conductance contributes to studies of surface chemistry, electrochemistry, and bonding dynamics.\cite{Xu2014, Li2012Charge, Kamenetska2009, Tao2008, Hla2003}

\begin{figure}[ht]%
	\includegraphics[width=\figurewidth,keepaspectratio,]{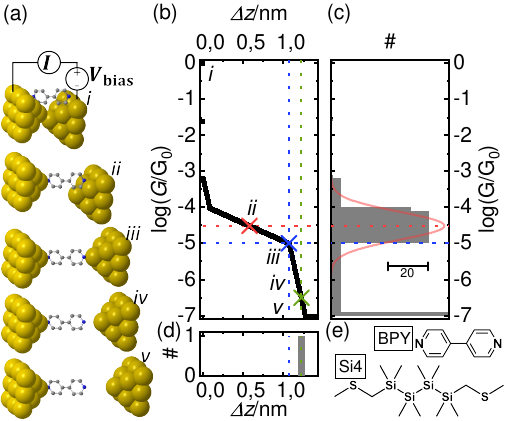}
	\caption{Idealized dummy trace and analysis. (a) An idealized break junction trajectory and (b) measured conductance trace following Au-Au rupture with (\textit{i}) Au-Au contact, (\textit{ii}) molecular plateau, (\textit{iii}) full extension, (\textit{iv}) return to tunneling, and (\textit{v}) noise floor. Red $\times$ marks customary $\overline{G}_{mol}$; blue $\times$ marks molecular plateau \acs{bp}; green $\times$ marks point in trace where customary plateau length is determined. (c) \acl{1DGH}, gray, with Gaussian fit to determine $\overline{G}_{mol}$ in (b). (d) This trace produces a count in a \acl{plh} at the green mark. (e) Molecular structures of \ac{bpy} and \ac{si4} studied in this report.}%
	\label{fig:dummy}%
\end{figure}

\Acp{smbj} are a leading method for measuring \acl{sm} conductance.
Fig.~\ref{fig:dummy}(a) is an idealized depiction of the evolution of a single \ac{smbj} and Fig.~\ref{fig:dummy}(b) depicts the resulting measured conductance trace.
In a \ac{smbj}, a molecular wire with anchoring ligands is introduced to a metal surface, either a flat metal substrate, or a metal wire.
This forms the first electrode and typically Au is the preferred metal (we will assume Au electrodes hereafter).
The second electrode, either a \ac{stm} tip, or the second half of the wire, is brought into contact with the first electrode (\textit{i}).
When the two electrodes are separated via a piezoelectric element, the Au-Au junction elongates and ruptures.
The elongated Au electrodes snap-back\cite{Huang2015break} yielding a drop in conductance much faster than the amplifier can track.
Once the amplifier catches up, the conductance follows a tunneling decay trend, unless a molecule incorporates into the junction.
In the case a molecule is incorporated, the conductance plateaus (\textit{ii}).
Ideally, the molecular junction reaches a fully-extended geometry before it ruptures (\textit{iii}).
After the molecular junction ruptures, the conductance trace returns to a tunneling decay trend (\textit{iv}) until the noise floor of the instrument is reached (\textit{v}).

\Ac{smbj} data records a range of conductance values as the electrodes are separated, as depicted in Figs.~\ref{fig:dummy}(a) and (b).
Conventionally a \acf{1DGH} [Fig.~\ref{fig:dummy}(c)] is calculated from a data set.
In the early years of \ac{smbj} experiments \acp{1DGH} were an important analysis tool to express molecular signals with low signal-to-noise ratios amid rather noisy data.\cite{Mayor2004, Xu2003Science}
Critically, the \ac{1DGH} contains the full range of conductance values from the trace and removes time and distance dependence from the data set.
This simplification was an acceptable transformation since the goal was to identify and compare a single unique mean molecular conductance value, $\overline{G}_{\text{mol}}$, for each target molecule.\cite{Salomon2003comparison}
The molecular conductance was modeled as a normally distributed conductance trace plateau whose variance was due to instrument noise and atomic rearrangements.\cite{Solomon2006, Xu2003JACS}
Viewing \acp{smbj} as a purely stochastic statistical problem where every new measure of molecular plateau is another noisy observation of the same underlying variable dictates that $\overline{G}_{\text{mol}}$ is the mean of all the plateaus in the data set, and a \ac{1DGH} is the correct tool for the job.
In a similar manner, a \ac{plh} [Fig.~\ref{fig:dummy}(d)] is often created by determining the extension of each trace at some conductance value after below the molecular conductance plateau -- a point necessarily after the junction ruptures.
Like with $\overline{G}_{\text{mol}}$, the mean molecular plateau length, $\overline{z}_{\text{mol}}$, is determined from a Gaussian fit to the \ac{plh}.
Information from the \ac{plh} is used to infer information about, for instance, relative ratios of different species measured throughout the experiment.\cite{Huang2017}

But if the molecular plateau is distance- and time-dependent, then lumping all the data points in the molecular plateau together into a \ac{1DGH}, data from points \textit{ii} and \textit{iii} in Fig.~\ref{fig:dummy}(a), will necessarily yield too-general of a value for $\overline{G}_{\text{mol}}$, and too long of a value for $\overline{z}_{\text{mol}}$.
What is needed, instead, is to choose a single point to label the molecular conductance and plateau length, and calculate the mean of those points only.
In this report we suggest that point \textit{iii} in Fig.~\ref{fig:dummy}(a), corresponding to the point in the junction trajectory when the molecule is the most extended in the junction immediately before junction rupture, is the best point to measure both the molecular conductance and molecular plateau length at.
In the conductance trace in Fig.~\ref{fig:dummy}(b), point \textit{iii} corresponds to the end of the molecular plateau, the \ac{bp}, before the conductance drops to the noise level of the instrument.
In this report we show that point \textit{iii} applies equally well to downward-sloping molecular plateaus typical of rigid molecular backbones, like \ac{bpy}, and to upward-sloping plateaus as observed in floppy molecules like alkanes and silanes [\acf{bpy} and \acf{si4} molecular structures are illustrated in Fig.~\ref{fig:dummy}(e)].

Generally speaking, finding point \textit{iii} in a molecular trace is a problem of estimating a structural break in a relationship between two variables (here, these two variables are distance and conductance).
A structural break marks the conductance at the point immediately before the Au-molecule-Au junction ruptures -- point \textit{iii} in Fig.~\ref{fig:dummy}(a) and (b).
It has been suggested that the last few \qty{}{\angstrom} of extension before the junction ruptures contains valuable information about the nature of the junction.\cite{Li2020}
For instance, previous studies have attempted to identify and characterize a final configuration of the break junction.\cite{Makk2012Pulling, balogh2015alternative}
In Ref.~\citenum{Makk2012Pulling}, the final configurations were identified by first using the rough estimate of plateau length, as described above, to mark the end of the plateau, and then comparing the final \qty{0.5}{\angstrom} of the conductance trace to the conductance histogram.
In this manner, the traces were classified into one of three final configurations.
Franco \latin{et al.} have further demonstrated that the early geometries of the junction provide little predictive power for how the junction ultimately elongates and ruptures.
The predictions of Franco \latin{et al.} provide strong impetus to focus experimentally on the region immediately prior to the structural break, and to focus analysis attention to this near-rupture region of the molecular trace.

A variety of interesting systems evolve smoothly under continuous force, yet at some point undergo a structural break and transition to a new local minimum, or to a new potential landscape, similarly to \acp{smbj}.
For instance, early models in finance tended to assume that stock prices move continuously via geometric Brownian motion, c.f. the famous Black and Scholes options pricing formula from the 1970s.\cite{Black1973}
However, some price movements seem better understood as discontinuous jumps –- for instance, in the “Flash Crash” of May 6th 2010, a large sell order in the Standard and Poor's 500 E-mini futures market led to a massive price decline and then reversal for many securities over a half hour period.\cite{Kirilenko2017}\bibnote{Over 300 securities saw prices decline over \qty{60}{\percent} before rebounding – see Ref.~\citenum{Kirilenko2017}}
Such events are consistent with the observed fat tails of stock return distributions, as described by Fama \latin{et al.}\cite{Fama1963}
And while financial models can take such discontinuous price movements into account (for options pricing see Ref.~\citenum{Merton1976}), to implement these in practice requires the ability to discern what qualifies as a jump.\cite{Merton1976}

Not surprisingly, economists have developed a mature toolbox for studying timeseries data that undergoes structural breaks -- a collection of algorithms targeted at \ac{cpd}.\cite{Truong2020}
Such breaks can take different forms, and have different economic interpretations.
For instance, is the shock a) temporary, and going to bounce back to the original trend in a V-shape (like the flash crash example)? Is it b) permanent, and a pure level shift that preserves the same slope and standard deviation (visually somewhat like a step function)?
In case a) the timeseries is trendstationary; while in case b) the timeseries exhibits a unit root.\cite{Donayre2021, Kwiatkowski1992, Elliott1992, PHILLIPS1988, Dickey1979}
On the other hand the structural break might signal a shift of the system into an entirely new state with new slope, standard deviation, and/or offset.

In this report we show that finding the structural break for each trace and averaging these \ac{bp} conductance values over the entire experiment reduces the standard deviation in the estimates for $\overline{G}_{\text{mol}}$.
But more importantly, this method is conceptually tailored to the specific problem of identifying $\overline{G}_{\text{mol}}$, and to comparing experimental results to theoretical calculations, whose models are simulating the geometry at point \textit{iii}.
When compared with other analysis methods, especially more recent \acl{ml} methods, the methodology reported in this report is transparent and simple, and it also eliminates the need for \latin{ad hoc} choosing of opaque \acl{ml} model parameters.
As demonstrated here, each trace has a single estimate for the structural \ac{bp}, and these points can be further studied as random samples from a distribution which can be summarized by $\overline{G}_{\text{mol}}$ and $\overline{z}_{\text{mol}}$, but also be used as feed-stock for more complicated analyses.
The critical improvement comes from a sharper focus on a more appropriate metric from the experiment.

\section{Methods}
\subsection{Experimental data}
All experimental results in this report were previously reported in the literature.
Interested readers can refer to Refs.~\citenum{Lauritzen2018, Magyarkuti2020, Su2017} for experimental methods therein.

\subsection{Change point detection with the Chow test}
The choice of \ac{cpd} algorithm is determined by the type of structural break expected in the timeseries.
Assumptions are made about trend-stationarity, unit roots, as well as numerous other behaviors like homo- or heterscedasticity, and long-range autocorrelated and nonlinear trends.\cite{Truong2020, Burg2020}
Nonetheless, all \ac{cpd} algorithms share three elements:
\begin{enumerate*}[(i)]
	\item constraint,
	\item search method, and
	\item cost function.
\end{enumerate*}

The constraint can simply be how many structural breaks are expected in the timeseries and can be a fixed integer.
In this report we limit our search to a single \ac{bp}, although many algorithms permit more than one \ac{bp}, and this constraint can be relaxed in the future.

The search method, too, can be as simple as inspecting each consecutive time point for the marker of a structural break.
This is the method applied in this report; however, in the future, more sophisticated search methods may be employed to reduce computational load and/or time.

The marker of a structural break is quantified by the cost function.
Numerous cost functions are proposed in the literature, accompanied by nuanced debate about the various merits.\cite{Truong2020}
One simple cost function is the Chow test.
The Chow test compares a single linear regression of the timeseries against the linear regressions of two segments created when the timeseries is divided at point $p$ (for a single \ac{bp}).
In other words, the timeseries of measurements, $Y$, with $N$ elements, measured at times, $X$, is divided at index $p$ into two linear regressions, left (L) and right (R),
\begin{equation}
	Y_{i=1,\dots,p} = a_L X_{i=1,\dots,p} + b_L + \varepsilon_L,
\end{equation}
consisting of the first part of the timeseries, running over indices $1, 2, ..., p$, and
\begin{equation}
	Y_{i=p+1,\dots,N} = a_R X_{i=p+1,\dots,N} + b_R + \varepsilon_R,
\end{equation}
consisting of the second part of the timeseries, running over indices $p+1, p+2, ..., N$.
The Chow test then calculates how much better the two linear regressions are than the single linear regression,
\begin{equation}
	Y_{i=1,\dots,N} = a_{\text{all}} X_{i=1,\dots,N} + b_{\text{all}} + \varepsilon_{\text{all}}.
\end{equation}
More specifically, the sum of the squared errors,\acused{rss}
\begin{equation}
	RSS_{L,R,all} = (\varepsilon_{L,R,all})^2,
\end{equation}
in these regressions are compared to $RSS_{all}$ from the error in the pooled regression, $\varepsilon_{\text{all}}$. For each $p$, the Chow score,
\begin{equation}\label{eq:chow}
	f(p) = \frac{\sfrac{[RSS_{\text{all}} - (RSS_1 + RSS_2)]}{k}}{\sfrac{(RSS_1 + RSS_2)}{(N - 2k)}}
\end{equation}
is calculated, where $k$ is the number of parameters to fit ($k=2$ in our case: $a$ and $b$). The index of the timeseries where the Chow score is greatest,
\begin{equation}\label{eq:bp}
	p_{\text{BP}} = \text{argmax}(f),
\end{equation}
is the index where $RSS$ of the two linear regressions is smallest, and therefore the best estimate for the \ac{bp}.

The Chow test is implemented in the package \textit{ruptures}, available as a Python package (https://centre-borelli.github.io/ruptures-docs/),\cite{Truong2018} but for our own understanding, we implemented our own subroutine for this report.

\subsection{Theoretical simulations to illustrate conductance-displacement trends in two target molecules}

We performed simulations of the molecular transport properties in this report using SIESTA and TranSIESTA.\cite{Soler2002, Brandbyge2002, Papior2017, Perdew1996}
Please refer to \ac{si} \S "Theoretical simulations to illustrate conductance-displacement trends in two target molecules" for details of the calculation.

\section{Results and Discussion \label{sec:results}}

\subsection{Chow test applied to idealized \ac{smbj} data to characterize the test}
\begin{figure}[]%
	\includegraphics[width=\figurewidth,keepaspectratio,]{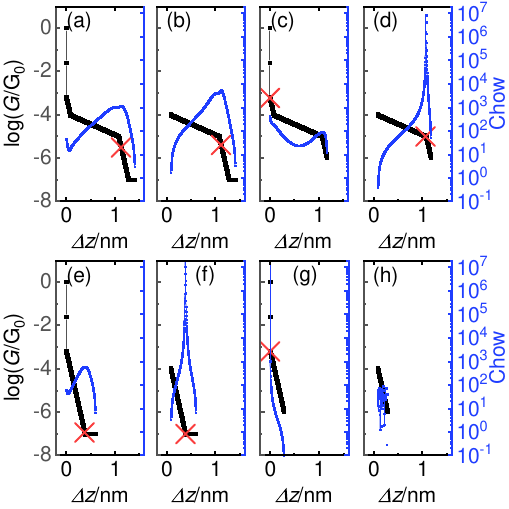}
	\caption{Chow score (blue) at each data point (black) applied to dummy \ac{smbj} molecular [(a)-(d)] and tunneling [(e)-(h)] traces with \ac{bp} (maximum of the Chow scores) marked with red $\times$. (a) and (e) full traces; (b) and (f) traces windowed from above; (c) and (g) traces windowed from below; and (d) and (h) traces windowed from both sides.}%
	\label{fig:dummy_chow}%
\end{figure}

We begin by showing the Chow test, Eqs.~\ref{eq:chow} and \ref{eq:bp}, applied to the idealized trace in Fig.~\ref{fig:dummy} to gain an intuition for how it behaves, starting with the whole trace (black) in Fig.~\ref{fig:dummy_chow}(a).
Interestingly, the Chow test (all scores in blue, \ac{bp} marked with red $\times$) identified a part of the trace following the \ac{bp} in Fig.~\ref{fig:dummy_chow}(a).
Next the trace was thresholded to focus the test below \qty{}{10^{-4}~\G} [Fig.~\ref{fig:dummy_chow}(b)], above \qty{}{10^{-6}~\G} [Fig.~\ref{fig:dummy_chow}(c)], and both [Fig.~\ref{fig:dummy_chow}(d)].
Removing the snap-back region [Fig.~\ref{fig:dummy_chow}(b)] had little effect on the result of the Chow test, but removing the noise data at the end of the trace [Fig.~\ref{fig:dummy_chow}(c)] had a large adverse effect.
Impressive results were achieved when the window was applied to both ends of the data [Fig.~\ref{fig:dummy_chow}(d)].
In this case, a cusp is achieved in the Chow scores, signaling a high confidence in the choice of \ac{bp}.
These tests suggested that, for most traces within a reasonable window, the Chow test will work well.

Blank traces are encountered in \ac{smbj} studies.
Typically, molecular experiments start with a number of junctions measured across bare Au-Au electrodes, before the molecule is introduced.
This is a way to establish a baseline, analogously to measuring a blank solvent in UV-vis.
Another common way to encounter blank traces is during the molecular experiment itself.
Depending on the junction formation probability of the molecular wire, the molecule may not bridge the two electrodes, and a blank trace is measured, even in a solution of molecules.
The junction formation probability is dependent on a number of factors including the stability of the instrument and the chosen anchor moieties and their preferred anchoring geometries, but it is still poorly understood, and it is an ongoing topic of investigation.\cite{Huang2015break, Kaliginedi2014promising}

At least a few blank traces are measured during a long experiment, even with analytes with high junction formation probabilities.
Therefore, it is important here to determine to what degree blank traces might confound any method based on \acp{bp}.

On the bottom row [Figs.~\ref{fig:dummy_chow}(e)-(h)] the same windows were applied to an idealized blank trace.
When no window was applied  [Fig.~\ref{fig:dummy_chow}(e)], the Chow test identified the inflection in the data when the trace reached the noise level, and likewise when the snap-back region was removed [Fig.~\ref{fig:dummy_chow}(f)].
When the noise level was removed [Fig.~\ref{fig:dummy_chow}(g)] the Chow test identified the inflection when the trace left the snap-back region.
In this case, too, a cusp signals a high confidence in choice of \ac{bp}; however, this \ac{bp} is in the noise floor.
Finally, when both windows were applied [Fig.~\ref{fig:dummy_chow}(h)] the Chow test had no confidence in a \ac{bp}.

The lesson here was that, with a reasonable window applied to each trace, which masks the noise floor and the snap back region, we can be confident the Chow test will yield a high score and accurately identify a \ac{bp} in a molecular trace, and yield no preferred value for a \ac{bp} in tunneling traces, or else a value in the noise floor.

\subsection{Chow test applied to a rigid molecular wire with a downward-sloping plateau}
The first experimental case we studied was \ac{bpy}, a well-studied molecular wire with a rigid backbone.\cite{Borges2016, Hamill2014Nanoscale, Perez-Jimenez2005JCP, Xu2003Science}
In this case, the molecular plateau was downward sloping, similar to the dummy data in Fig.~\ref{fig:dummy}.
A downward sloping molecular plateau will yield an over-abundance of data points higher in conductance than the molecule in its final, extended geometry, and result in an over-estimate of $\overline{G}_{\text{mol}}$.
Moreover, \ac{bpy} is known to exhibit two molecular plateaus at room temperature, possibly causing an even wider spread in the distribution of molecular-region data points.
We studied two \ac{bpy} data sets, one at cryogenic temperature, and one at \acl{rt}.
It was expected that \qty{4}{\K} data would have sharp \acp{bp} and achieve good results applying the Chow test. Comparing \qty{4}{\K} with \acl{rt} data allowed us to make a comparison between "easy" data and "hard" data.

\subsubsection{Simulation of \ac{bpy} conductance-displacement dependence}
The bonding angle-dependence of \ac{bpy} has been well studied theoretically and experimentally.\cite{Makk2012, Quek2009NN}
In Ref.~\citenum{Quek2009NN} it was reported that the conductance drops by an order of magnitude as the bonding angle goes from \qtyrange{50}{90}{\degree}, where \qty{90}{\degree} is perpendicular, in theoretical transport calculations.
Another way to say this is that, any single \ac{bpy} molecular junction will likely sample geometries which yield conductance values with variance of approximately an order of magnitude.
Furthermore, the values which are measured near the beginning of the plateau, which likely correspond to bonding angles of \qty{\approx50}{\degree}, will be approximately one order of magnitude higher in conductance than the values which are measured near the end of plateau, where the bonding angle is likely closer to \qty{\approx90}{\degree}.
Including the data points at the beginning of the plateau in fit to determine for $\overline{G}_{\text{mol}}$ will not accurately estimate $\overline{G}_{\text{mol}}$, due to the distance-dependence of the conductance, and not correspond to a simulation, if those transport calculations use geometries of an extended junction with bonding angle of \qty{90}{\degree}.

\begin{figure}[]%
	\includegraphics[width=\figurewidth,keepaspectratio,]{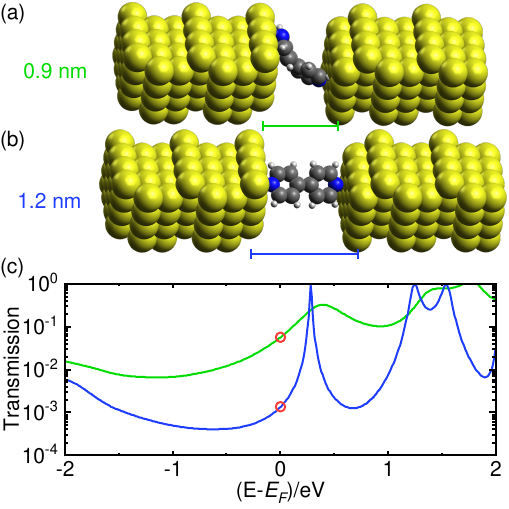}
	\caption{Geometries for transport calculations of (a) contracted and (b) extended \ac{bpy} junctions and estimated electrode-to-electrode displacements; (c) transmission versus energy curves for contracted (green) and extended (blue) geometries from (a) and (b), respectively, with transmission at the Fermi energy emphasized with a red $\bigcirc$.}%
	\label{fig:BPY_sim}%
\end{figure}

To illustrate this, we simulated the transmission of \ac{bpy} in junctions with two different separations of the electrodes.
In the geometry in Fig.~\ref{fig:BPY_sim}(a), the electrodes are too close (\qty{0.9}{\nm}) to fit the extended molecule, so a tilt angle is necessary for the \ac{bpy} to fit in the junction.
In the geometry in Fig.~\ref{fig:BPY_sim}(b), the electrodes are separated by a distance (\qty{1.2}{\nm}) which permits the \ac{bpy} to fit extended in the junction.
As reported previously,\cite{Quek2009NN} the calculated transmission drops by over an order of magnitude between these two geometries, as shown in Fig.~\ref{fig:BPY_sim}(c).
This conductance change happens as the displacement changes by only \qty{0.3}{\nm} in our simulation.

\subsubsection{\ac{bpy} at cryogenic temperature, when trace features are sharp}
\begin{figure}[]%
	\includegraphics[width=\figurewidth,keepaspectratio,]{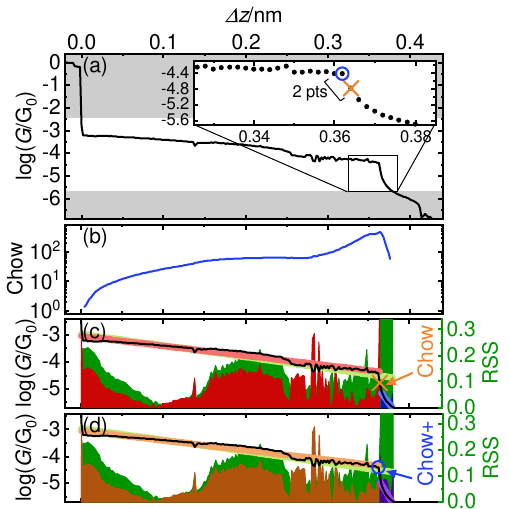}
	\caption{Chow test applied to an example trace of \ac{bpy} measured at \qty{4}{\K}. (a) example conductance vs displacement trace; inset zoom to end of molecular plateau with \ac{bp} determined by Chow test (orange $\times$) and Chow+ (blue $\bigcirc$); Chow+ searches \num{2} data points before the \ac{bp} chosen by the Chow test for the largest jump in conductance to determine an improved \ac{bp}. (b) Chow score (blue) vs plateau data point for windowed region of trace. (c) linear regressions (pooled data in green, segment before \ac{bp} in red, and segment after \ac{bp} in blue) and \ac{rss} at \ac{bp} determined by Chow test. (d) linear regressions (pooled data in green, segment before \ac{bp} in orange, and segment after \ac{bp} in purple) and \ac{rss} at \ac{bp} determined by Chow+.}%
	\label{fig:trace}%
\end{figure}

Fig.~\ref{fig:trace}(a) plots a single example trace from a data set of \num{5053} traces of \ac{bpy} as analyte, of which \num{1843} traces contained clear molecular features, permitting hand-sorting the traces into blank and molecular traces.
The data and methods were previously reported in Refs.~\citenum{Lauritzen2018} and \citenum{Magyarkuti2020}.
Based on the conclusions from the discussion around Fig.~\ref{fig:dummy_chow}, the trace was masked [gray regions in Fig.~\ref{fig:dummy_chow}(a)] to focus on conductances between \qty{}{10^{-2.4}~\G} and \qty{}{10^{-5.7}~\G}. Fig.~\ref{fig:trace}(b) plotted the Chow score tested at each data point in the range.
The sensitivity of these methods to the window parameters is discussed in further detail in \ac{si} \S "Sensitivity testing of four parameters used in the Chow+ test."
The maximum Chow score, orange $\times$ in Fig.~\ref{fig:trace}(c), was slightly below the expected \ac{bp}.
Fig.~\ref{fig:trace}(c) plotted the \ac{rss} for the entire trace segment (linear regression in light green and \ac{rss} in dark green), and the segment before/after the Chow score maximum (linear regression in light red/blue, and \ac{rss} in dark red/blue).

The lesson from Fig.~\ref{fig:dummy_chow} was that the exact choice of \ac{bp} determined by the Chow test may be skewed to lower conductances, likely a result of the particular search window.
Because it was inefficient to choose a unique window for each trace, but instead one window for the entire data set, this lower skewing was predicted to alter the results of the Chow test method for determining the \ac{bp} for an entire data set.
Another way to put this, the Chow test was a good and fast approach for searching globally for a region where the \ac{bp} was located. But to fine-tune our result, we implemented a second step to the search.
In this second step, we searched \num{2} data points before the \ac{bp} determined by the Chow test for the largest jump in the data, and chose the left-most data point in that jump.
The combined \ac{bp} search, first using the Chow test to identify an approximate \ac{bp}, and next using the search back, we called Chow+.
The inset in Fig.~\ref{fig:trace}(a) depicts such a search.
The orange $\times$ marks the choice based solely on the Chow test; the blue $\bigcirc$ marks the choice after searching for the largest jump in the previous \num{2} data points.
For all data sets in this report, the search back distance was always \num{2} data points.
The sensitivity of these results to the number of points we searched back is discussed further in \ac{si} \S "Sensitivity testing of four parameters used in the Chow+ test."
Fig.~\ref{fig:trace}(d) plotted the linear regressions and \ac{rss} for the new segments created by the new choice of \ac{bp}.
The difference, in this case, was one data point, and there was little difference in the \ac{rss} between Figs.~\ref{fig:trace}(c) and (d).
Fig.~\ref{fig:trace}(b), too, showed there was little difference between the Chow scores of the two points. Furthermore, the change in displacement between the two methods, one data point, was \qty{2.0}{\pm}.
Yet, the change in conductance was not negligible: a change of \qty{1.7}{\nano\siemens}.

\begin{figure}[]%
	\includegraphics[width=\figurewidth,keepaspectratio,]{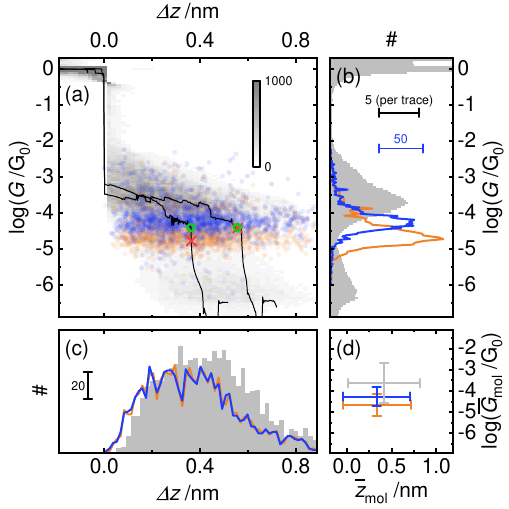}
	\caption{Chow test applied to \num{1843} traces of \ac{bpy} at \qty{4}{\K}. (a) \ac{2DH} of all molecular traces in gray; scatter plots of Chow/Chow+ \acp{bp} in orange/blue; two example traces with Chow/Chow+ \acp{bp} plotted in red/green. (b) \ac{1DGH} of all traces (gray), and conductance values of Chow/Chow+ test results (orange/blue). (c) \Ac{plh} (gray), and histograms of displacement values of Chow/Chow+ test results (orange/blue). (d) Mean of Gaussian fit to \ac{1DGH} of all data between \qty{}{10^{-1.0}~\G} and \qty{}{10^{-4.9}~\G} (gray) and \acl{plh}, and to conductances and displacements of Chow/Chow+ test results (orange/blue). Error bars are standard deviations.}%
	\label{fig:BPY_cold}%
\end{figure}

The \acp{bp} for all the molecular traces in the data set were then calculated, plotting both the simple Chow test (Chow, orange points) results and the Chow test plus correction (Chow+, blue points) results in Fig.~\ref{fig:BPY_cold}(a).
A \ac{2DH} of only molecular traces was plotted in gray in Fig.~\ref{fig:BPY_cold}(a).
The trace used in Fig.~\ref{fig:trace} was also plotted, with a red $\times$ at the Chow result, and a green $\bigcirc$ at the Chow+ result. The Chow results, without correction, yielded, on average, results much lower in conductance than where the molecular plateau appeared to end, but instead were located halfway down the break between the molecular plateau and the beginning of the noise level.
On the other hand, the Chow+ results, on average, appeared to be located just following the end of the molecular plateau.
When a histogram of the conductance of the \acp{bp} was plotted along with the \ac{1DGH} of the data set in Fig.~\ref{fig:BPY_cold}(d), the difference between $\overline{G}_{\text{mol}}$ from the Chow+ \acp{bp} and $\overline{G}_{\text{mol}}$ from the \ac{1DGH} was over a half order magnitude, a change of \qty{14.0}{\nano\siemens}.
$\overline{z}_{\text{mol}}$ changed only \qty{0.9}{\angstrom} between the conventional \acl{plh} and a histogram created from the displacements of the \acp{bp}, plotted in Fig.~\ref{fig:BPY_cold}(c).
$\overline{G}_{\text{mol}}$ and $\overline{z}_{\text{mol}}$, as determined by Gaussian fits of the \ac{1DGH} and \ac{plh}, and from the Chow and Chow+ tests were plotted in Fig.~\ref{fig:BPY_cold}(d), with the standard deviations used for error bars.
It is noteworthy that the Chow+ approach reduced the standard deviation of the conductance by half, compared to the conventional \ac{1DGH} approach, meaning the accuracy with which $\overline{G}_{\text{mol}}$ is known was improved using Chow+ over the conventional method.
Consequently, we applied the Chow+ test throughout the remainder of this study.

\begin{figure}[]%
	\includegraphics[width=\figurewidth,keepaspectratio,]{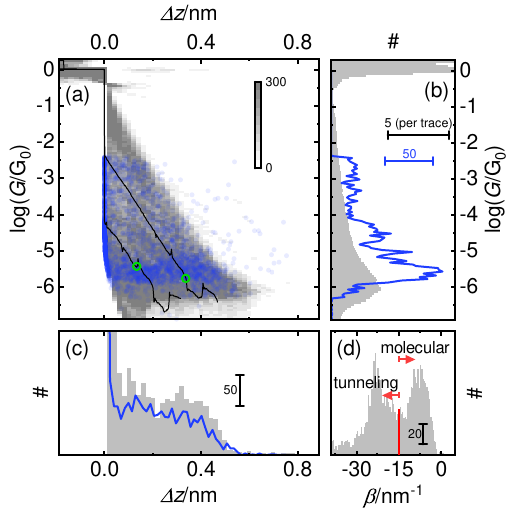}
	\caption{Chow+ test applied to \num{3210} tunneling traces at \qty{4}{\K}. (a) \ac{2DH} of all molecular traces in gray; scatter plots of Chow+ \acp{bp} in blue; two example traces with Chow+ \acp{bp} plotted in green. (b) \ac{1DGH} of all tunneling traces (gray), and conductance values of Chow+ test results (blue). (c) \acl{plh} (gray), and histograms of displacement values of Chow+ test results  (blue). (d) Histogram of slopes of each trace \qty{1}{\angstrom} previous to \ac{bp} (gray) for the entire data set, both molecular and blank traces. Red line marks threshold between mostly tunneling traces and mostly molecular traces.}%
	\label{fig:tunnel_cold}%
\end{figure}

Most \ac{smbj} experiments contain at least some tunneling traces which result when a molecule was not captured in the \acl{bj}.
The complete data set above contained \num{3210} traces which contained little or no molecular signal, as determined by hand-selection.
To confirm the distribution of \acp{bp} in Fig.~\ref{fig:BPY_cold} was unique to the molecule, the tunneling traces were analyzed in an identical manner.
A \ac{2DH} of the tunneling traces, overlayed with a scatter plot of \acp{bp}, determined with the Chow+ test, were plotted in Fig.~\ref{fig:tunnel_cold}(a).
There was an intense region around \qty{}{10^{-5.5}~\G}, as confirmed by a histogram of conductance values, Fig.~\ref{fig:tunnel_cold}(b), which were likely marking the inflection of the trace from tunneling to noise, identical to the result on the dummy trace in Figs.~\ref{fig:dummy_chow}(e) and (f).
This intense region was below the results for the molecular data in Figs.~\ref{fig:BPY_cold}(a) and (b).
Indeed, there appeared to be some \acp{bp} with tunneling nature in the molecular data.
A histogram of displacements, Fig.~\ref{fig:tunnel_cold}(c), helped to illustrate the distribution of displacements in the \acp{bp}.

Here it is important to note, there are a number of benefits of finding the \ac{bp}.
One example is the increased resolution in the slope of the trace, which in turn contains information about the time-dependence of the \acl{bj} and the presence or absence of the molecule, but countless other analyses might be performed once the \ac{bp} is identified.
In this study, once the \ac{bp} of the trace was determined using the Chow+ test, the slope of each trace was determined by taking the slope of the previous \qty{1}{\angstrom} of the trace.
This increased the accuracy over conventional windowing methods by only using data that is in the molecular trace without contamination from the steep portions following the Au-Au rupture or \ac{bp}.
The increased accuracy allowed us to observe a sharp bimodal distribution in the slopes when all traces were used, both molecular and tunneling, to make a histogram of slopes [Fig.~\ref{fig:tunnel_cold}(d)]. This bimodal distribution proved to be an effective tool to sort the data into tunneling traces with steep slopes, and molecular traces with much flatter slopes, although to remain consistent with Refs.~\citenum{Lauritzen2018} and \citenum{Magyarkuti2020}, we used the hand labeling employed therein for labeling in this report.

In Ref.~\citenum{Li2020} Li \latin{et al.} argued that the last few \qty{}{\angstrom} of the junction were independent of the geometry of the Au-Au junction that preceded it, in simulated junctions, and the fluctuations were assigned entirely to thermal fluctuations.
Thus, the last \qty{1}{\angstrom} before the \ac{bp} contains the most accurate experimental representation of a molecule trapped between two electrodes, and vibrating solely due to thermal excitations, and this moment is best identified by using the \ac{bp} to determine it.

\subsubsection{\ac{bpy} at \acl{rt}, when thermal fluctuations influence trace features}
\begin{figure}[]%
	\includegraphics[width=\figurewidth,keepaspectratio,]{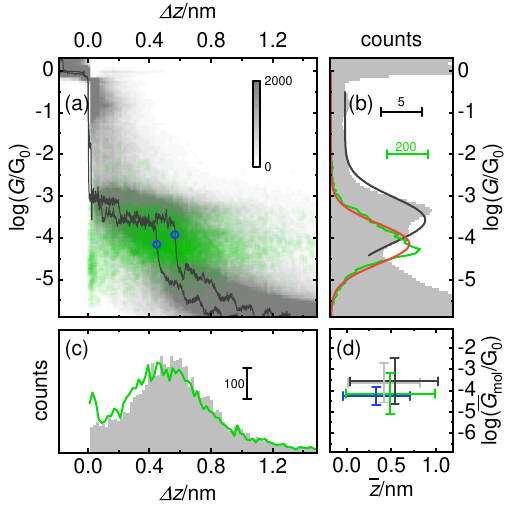}
	\caption{Chow test applied to \num{9847} traces of \ac{bpy} at \acl{rt}. (a) \ac{2DH} of all molecular traces in gray; scatter plots of \acp{bp} in green; two example traces with \acp{bp} plotted in blue.  (b) \ac{1DGH} of all traces (gray), and  histogram of conductance values of Chow+ test results (green). Gaussian fits to \ac{1DGH} (dark gray) and \acp{bp} (red) included as aid for the eye. (c) \acl{plh} (gray), and histograms of displacement values of Chow+ test results (green). (d) Mean of Gaussian fit to \ac{1DGH} of all data between \qty{}{10^{-0.4}~\G} and \qty{}{10^{-4.4}~\G} (dark gray), and to conductances and displacements of Chow+ test results (green). Results for \qty{4}{\K} measurements (light gray and blue) included for comparison. Error bars are standard deviations.}%
	\label{fig:BPY_rt}%
\end{figure}

Next the Chow+ test was applied to a data set of \ac{smbj} measurements on \ac{bpy} at \acl{rt}. Fig.~\ref{fig:BPY_rt}(a) shows a \ac{2DH} of all traces in the data set, with \acp{bp} plotted in green.
Two example traces are also plotted, with blue $\bigcirc$s to show the Chow+ test choice of \ac{bp}.

Due to the particular nature of this data, a Savitzky-Golay smoothing filter\cite{Savitzky1964} (as implemented in SciPy\cite{2020SciPy_NMeth}) with polynomial order \num{2} and smoothing window of \num{51} data points was necessary before the Chow test was applied.
As with the \qty{4}{\kelvin} data, the snap-back and tunneling portions of each trace were masked.
The choice for this window is discussed in further detail in \ac{si} \S "Sensitivity testing of four parameters used in the Chow+ test."

The choices of \ac{bp} appeared to be accurate, as well, in the \acl{rt} \ac{bpy} data, based on the expected results in the example traces.
The distribution of \acp{bp} was more clearly seen in histograms of conductance [Fig.~\ref{fig:BPY_rt}(b)] and displacements [Fig.~\ref{fig:BPY_rt}(c)].
It is well known \ac{bpy} at \acl{rt} exhibits a 2-step feature during junction elongation, and the conventional histograms in Figs.~\ref{fig:BPY_rt}(a) and (b) showed these features, while the Chow+ result was mostly insensitive to this behavior.
This is a strength of using Chow+.
Various molecules, e.g. \ac{bpy} here or the \ac{si4} in the next section, are expected to exhibit a wide range of junction trajectories which may confound accurate estimates of $\overline{G}_{\text{mol}}$ and $\overline{z}_{\text{mol}}$.
But eventually the molecular junction breaks, and this is likely to be the best estimate for $\overline{G}_{\text{mol}}$ and $\overline{z}_{\text{mol}}$.
Fig.~\ref{fig:BPY_rt}(d) summarizes the estimates for $\overline{G}_{\text{mol}}$ and $\overline{z}_{\text{mol}}$, as determined by conventional \ac{1DGH}, \acl{plh}, and Chow+, for \ac{bpy} at \qty{4}{\K} [light gray and blue, from Fig.~\ref{fig:BPY_cold}(d)] and \acl{rt} (dark gray and green).
The differences between Chow+ and conventional estimates for $\overline{G}_{\text{mol}}$ and $\overline{z}_{\text{mol}}$ in the \acl{rt} data were \qty{15.3}{\nano\siemens} and \qty{0.5}{\angstrom}.
The standard deviation in the \ac{bp} conductance and displacement distributions were not significantly different than the conventional histograms, and no improvement was made in the error in these estimates.
However, the estimate for $\overline{G}_{\text{mol}}$ was over a half-order of magnitude lower in conductance than the estimate from the \ac{1DGH}.

It is also noteworthy that the Chow+ test finds estimates which are shorter, and with lower conductance, than the conventional methods.
This is not the case in the next example.

\subsection{Chow test applied to a floppy molecular wire with an upward-sloping plateau}
The Chow+ identified a $\overline{G}_{\text{mol}}$ in \ac{bpy}, which has a downward sloping molecular plateau with steps, that was lower in conductance than the customary fit to the \ac{1DGH} peak.
This was the case for both cryogenic data with sharp features, and \acl{rt} data with thermally influenced features.
We next applied Chow+ to \ac{si4} at \acl{rt}.

\subsubsection{Simulation of \ac{si4} conductance-displacement dependence}

For \ac{si4}, the conductance-displacement dependence is different than for \ac{bpy}.
Ref.~\citenum{Su2017} reported a stereoelectric relationship between the methyl anchoring geometry and the conductance to explain the rise during junction elongation.
Another possibility is similar to that reported for alkanes\cite{Fujihira2006}: when the electrodes are closer than the length of the molecule, at small displacements, gauche defects permit the molecule to fit in the junction, but yield lower conductance.
These defects are relaxed as the junction elongates, resulting in a rise in conductance.

\begin{figure}[]%
	\includegraphics[width=\figurewidth,keepaspectratio,]{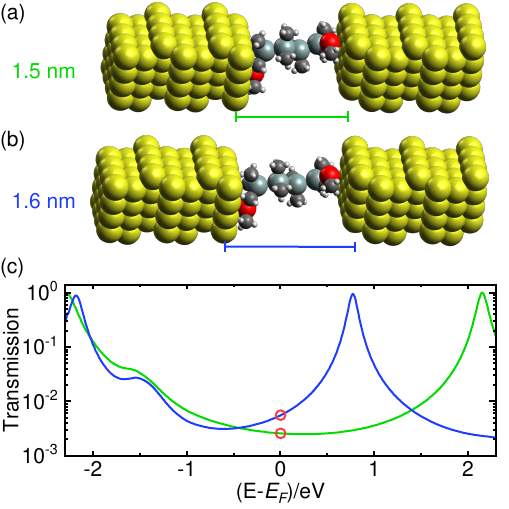}
	\caption{Geometries for transport calculations of (a) contracted and (b) extended \ac{si4} junctions and estimated electrode-to-electrode displacements; (c) transmission versus energy curves for contracted (green) and extended (blue) geometries from (a) and (b), respectively, with transmission at the Fermi energy emphasized with a red $\bigcirc$.}%
	\label{fig:Si4_sim}%
\end{figure}

To illustrate this, we calculated the transport of \ac{si4} at two different separations of the electrodes.
In the smaller separation, Fig.~\ref{fig:Si4_sim}(a), there was not enough room for the fully extended \ac{si4} molecule (\qty{1.5}{\nm}), so a gauche defect was imposed to make it fit.
The larger separation, Fig.~\ref{fig:Si4_sim}(b), was chosen to allow the fully extended molecule to fit (\qty{1.6}{\nm}).
The difference in the transmission between these two geometries was approximately a half order magnitude, as shown in Fig.~\ref{fig:Si4_sim}(c).
Significantly, the smaller separation yielded a lower transmission than the larger separation.

\subsubsection{Chow+ test applied to experimental \ac{si4} traces}

\begin{figure}[]%
	\includegraphics[width=\figurewidth,keepaspectratio,]{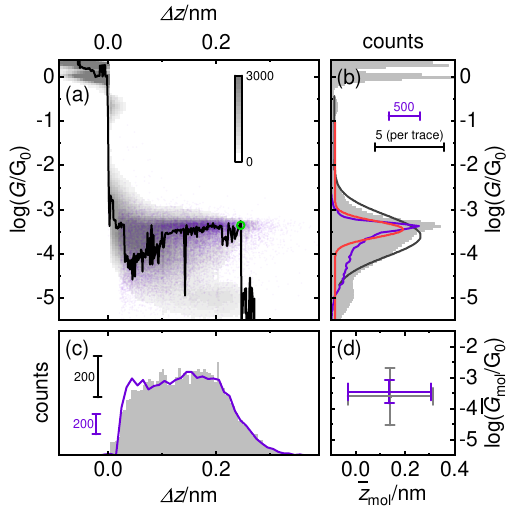}
	\caption{Chow test applied to \num{14119} traces of \ac{si4} at \acl{rt}. (a) \ac{2DH} of all molecular traces in gray; scatter plots of \acp{bp} in purple; example trace with \acp{bp} plotted in green. (b) \ac{1DGH} of all traces (gray), and  histogram of conductance values of Chow+ test results in purple. Gaussian fits to \ac{1DGH} (dark gray) and \acp{bp} (red) included as aid for the eye. (c) \Ac{plh} (gray), and histograms of displacement values of Chow+ test results (purple). (d) Mean of Gaussian fit to \ac{1DGH} of all data between \qty{}{10^{-0.5}~\G} and \qty{}{10^{-4.5}~\G}  and \acl{plh} (gray), and to conductances and displacements of Chow+ test results (purple). Error bars are standard deviations.}%
	\label{fig:Si}%
\end{figure}

The \ac{si4} data set contained \num{20189} traces, from which \num{14119} traces were identified as molecular traces, determined by the slope of the last \qty{0.2}{\angstrom} of the plateau before the \ac{bp}.
This short distance to determine the slope was chosen because the plateaus were very short, and because the sampling rate was relatively high, permitting reasonably good statistics.
Fig.~\ref{fig:Si} plotted the results of the molecular traces. The data were previously reported in Ref.~\citenum{Su2017}.
The frequent and large jumps in conductance were previously attributed to a stereoelectric effect due to the methyl anchoring geometry.
For the purposes of the present study, these jumps present a potential confounding feature in the data for our simplistic \ac{cpd}.
Besides this, these jumps also spread the distribution of conductance values in the molecular region, needlessly increasing the uncertainty of $\overline{G}_{\text{mol}}$, when this value is determined by a Gaussian fit to the \ac{1DGH}.
Therefore, \ac{si4} data provides an opportunity both to demonstrate the success of the Chow+ method in the face of one type of potentially confounding trace features, and also to demonstrate the success of using \acp{bp} to significantly reduce the uncertainty in determining $\overline{G}_{\text{mol}}$, for the positive benefit of any downstream analysis.

Due to the high sampling rate in this data set, and the abundance of large jumps in the plateaus, a Savitzky-Golay smoothing filter with polynomial order \num{2} and window of \num{45} points was applied.
This smoothing, along with a search back of \num{2} data points after the Chow test was applied, yielded satisfactory results.
As before, each trace was windowed to remove the snap-back and noise floor portions.
The sensitivity of the following analysis to these choices of parameters is discussed further in \ac{si} \S "Sensitivity testing of four parameters used in the Chow+ test."

Fig.~\ref{fig:Si}(a) showed \acp{bp} mostly distributed tightly along the part of the \ac{2DH} where the majority of traces had reached a plateau, after any drop and rise of the conductance earlier in the plateau.
The example trace in Fig.~\ref{fig:Si}(a) was representative of this -- the \ac{bp} marked the end of the flat part of the trace.
\Acp{bp} chosen at this point were likely to represent fully extended junctions, comparable to geometries used by theory to model the junction. Conversely, values for $\overline{G}_{\text{mol}}$ derived from the \ac{1DGH}, Fig.~\ref{fig:Si}(b), will necessarily contain nature from the conductance jumps -- geometries other than straight elongated junctions.
The narrowing of the conductance distribution between the conventional method (gray) and Chow+ (purple) was evident in Fig.~\ref{fig:Si}(b), though the distribution of displacements, Fig.~\ref{fig:Si}(c) appeared to change little.
The summary in Fig.~\ref{fig:Si}(d) for $\overline{G}_{\text{mol}}$ and $\overline{z}_{\text{mol}}$ showed the standard deviation for the estimate of $\overline{G}_{\text{mol}}$ was reduced by a third, from \qty{650}{\nano\siemens} to \qty{180}{\nano\siemens}, between the conventional method and Chow+.

\begin{figure}[]%
	\includegraphics[width=\figurewidth,keepaspectratio,]{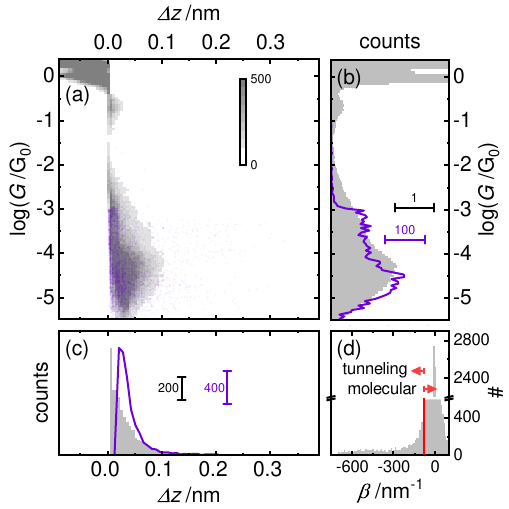}
	\caption{(a) \ac{2DH} (gray) and scatter plot of \acp{bp} (purple), (b) \ac{1DGH} (gray) and histogram of \ac{bp} conductances (purple), and (c) \ac{plh} (gray) and histogram of \ac{bp} displacements of tunneling traces sorted out of data set of \ac{si4} \acp{bj}. (d) Histograms of slopes of all traces in the data set; slopes were determined from the last \qty{1}{\angstrom} of the molecular plateau before the \ac{bp} as determined by Chow+. Red line in (d) marks the threshold for labeling: traces with slopes greater than \qty{-80}{\per \nm} were labeled "tunneling", less than \qty{-80}{\per \nm} were labeled "molecular."}%
	\label{fig:Si_tun}%
\end{figure}

For this data set it proved necessary to separate out blank traces to accurately estimate $\overline{G}_{\text{mol}}$ only from the molecular traces.
The tunneling data is summarized in Fig.~\ref{fig:Si_tun}, which shows little presence of molecular conductance plateaus in the \ac{2DH} [gray in Fig.~\ref{fig:Si_tun}(a)] or molecular conductance peaks in the \ac{1DGH} [gray in Fig.~\ref{fig:Si_tun}(b)].
Likewise, the histograms of the \ac{bp} conductances [purple in Fig.~\ref{fig:Si_tun}(a)] shows no molecular peaks.
The \ac{plh} [gray in Fig.~\ref{fig:Si_tun}(c)] shows plateau lengths much shorter than the molecular plateaus, and the histogram of \ac{bp} displacements also contains only sort \acp{bp}.
These sorting results were achieved by determining the slope of the last \qty{0.2}{\angstrom} for each trace.
A histogram of these slopes is shown in Fig.~\ref{fig:Si_tun}(d).
The sorting between the molecular traces in Fig.~\ref{fig:Si} and the tunneling traces in Fig.~\ref{fig:Si_tun} was achieved by a simple threshold at \qty{-80}{\per \nm}, where tunneling traces had slopes larger than the threshold, and molecular traces had slopes smaller than the threshold.

\section{Conclusion}
We have presented a proof-of-concept for applying \ac{cpd} to determine the molecular conductance and displacement of a \ac{smbj} trace. In every example we have analyzed, the resulting summary statistics provide estimates for $\overline{G}_{\text{mol}}$ with smaller standard deviation, typically by half, compared to the conventional \ac{1DGH} method.
These results are achieved despite explicit confounding features in the data, including noise due to sample rates and electronic sources, and molecular-specific sources like two-step features and dips-and-rises.

In this study we utilized the Chow test because of its simplicity.
The Chow test has a couple of shortcomings; its conventional implementation is slow, and it assumes as the null hypothesis that a single unit root exists, even if there are more than one, or none at all.
In most \ac{smbj} traces this is a safe assumption, and we have presented results on molecules which exhibit more than one apparent \acp{bp} without excessive complications.
In the case of multiple \acp{bp}, other \ac{cpd} algorithms may be better suited.
Most notably, advanced \acl{ml} methods were not considered here.
One reason for this choice was that we wished to present a counterpoint to \acl{ml}, based on the premise that, when faced with an intractable problem, most researchers do not need more complicated \acl{ml}, only better statistics.
Contrary to most, or all, \acl{ml} methods, Chow+ requires four physically meaningful tuning parameters, three of which (window ceiling and floor, and smoothing window) are typically applied during \acl{ml} workflow as well.
The fourth, the search back distance, is \latin{ad hoc}, but does not need to be used at all, and when it is used, provides only small corrections to the algorithm and sensible improvements to the results.

The strengths of applying \ac{cpd} to determine $\overline{G}_{\text{mol}}$ and $\overline{z}_{\text{mol}}$ are manifold.
Determining a single scalar value from each trace will help with \acl{ml} approaches, which are currently searching around for effective dimensionality reduction devices.
A single scalar value from each trace can be plotted in a timeseries, providing researchers with a tool to monitor experiments online, and poll real-time statistics.
With improvements, methods to search for multiple \acp{bp} will be useful for studying in further detail molecules like \ac{bpy} and \ac{si4} with clear and explainable non-linear behaviors in the molecular plateaus.
In light of reports by Franco \latin{et al.}, we are confident that focusing analysis attention on the part of the molecular trace immediately preceding the \ac{bp} will prove fruitful and easier than using the entire molecular plateau.
Finally, we expect that the symbiosis of experiment and theory will improve when molecular conductances derived from \ac{bp} analysis are used instead of \acp{1DGH}.

\begin{acknowledgement}
We thank Professors Timothy Su and Latha Venkataraman for sharing data and their valuable feedback.
This project has received funding from the European Research Council (ERC) under the European Union’s Horizon 2020 research and innovation programme grant agreement No 865870 and Marie Skłodowska-Curie grant agreement No 884741.
\end{acknowledgement}

\begin{suppinfo}
The following files are available free of charge.
\begin{itemize}
	\item Supporting-Information.pdf: Details of theoretical calculations; details of sensitivity testing for the four necessary parameters for the Chow+ test.
\end{itemize}
\end{suppinfo}

\bibliography{library-break_point_v1-10-acs.bib}

\providecommand{\latin}[1]{#1}
\makeatletter
\providecommand{\doi}
  {\begingroup\let\do\@makeother\dospecials
  \catcode`\{=1 \catcode`\}=2 \doi@aux}
\providecommand{\doi@aux}[1]{\endgroup\texttt{#1}}
\makeatother
\providecommand*\mcitethebibliography{\thebibliography}
\csname @ifundefined\endcsname{endmcitethebibliography}
  {\let\endmcitethebibliography\endthebibliography}{}
\begin{mcitethebibliography}{56}
\providecommand*\natexlab[1]{#1}
\providecommand*\mciteSetBstSublistMode[1]{}
\providecommand*\mciteSetBstMaxWidthForm[2]{}
\providecommand*\mciteBstWouldAddEndPuncttrue
  {\def\EndOfBibitem{\unskip.}}
\providecommand*\mciteBstWouldAddEndPunctfalse
  {\let\EndOfBibitem\relax}
\providecommand*\mciteSetBstMidEndSepPunct[3]{}
\providecommand*\mciteSetBstSublistLabelBeginEnd[3]{}
\providecommand*\EndOfBibitem{}
\mciteSetBstSublistMode{f}
\mciteSetBstMaxWidthForm{subitem}{(\alph{mcitesubitemcount})}
\mciteSetBstSublistLabelBeginEnd
  {\mcitemaxwidthsubitemform\space}
  {\relax}
  {\relax}

\bibitem[Veselinovic \latin{et~al.}(2019)Veselinovic, Alangari, Li, Matharu,
  Art{\'{e}}s, Seker, and Hihath]{Veselinovic2019}
Veselinovic,~J.; Alangari,~M.; Li,~Y.; Matharu,~Z.; Art{\'{e}}s,~J.~M.;
  Seker,~E.; Hihath,~J. Two-tiered electrical detection, purification, and
  identification of nucleic acids in complex media. \emph{Electrochimica Acta}
  \textbf{2019}, \emph{313}, 116--121\relax
\mciteBstWouldAddEndPuncttrue
\mciteSetBstMidEndSepPunct{\mcitedefaultmidpunct}
{\mcitedefaultendpunct}{\mcitedefaultseppunct}\relax
\EndOfBibitem
\bibitem[Loh \latin{et~al.}(2018)Loh, Burgess, Tanase, Ferrari, McLachlan,
  Cass, and Albrecht]{Loh2018}
Loh,~A. Y.~Y.; Burgess,~C.~H.; Tanase,~D.~A.; Ferrari,~G.; McLachlan,~M.~A.;
  Cass,~A. E.~G.; Albrecht,~T. Electric Single-Molecule Hybridization Detector
  for Short {DNA} Fragments. \emph{Analytical Chemistry} \textbf{2018},
  \emph{90}, 14063--14071\relax
\mciteBstWouldAddEndPuncttrue
\mciteSetBstMidEndSepPunct{\mcitedefaultmidpunct}
{\mcitedefaultendpunct}{\mcitedefaultseppunct}\relax
\EndOfBibitem
\bibitem[Storm \latin{et~al.}(2003)Storm, Chen, Ling, Zandbergen, and
  Dekker]{Storm2003}
Storm,~A.~J.; Chen,~J.~H.; Ling,~X.~S.; Zandbergen,~H.~W.; Dekker,~C.
  Fabrication of solid-state nanopores with single-nanometre precision.
  \emph{Nature Materials} \textbf{2003}, \emph{2}, 537--540\relax
\mciteBstWouldAddEndPuncttrue
\mciteSetBstMidEndSepPunct{\mcitedefaultmidpunct}
{\mcitedefaultendpunct}{\mcitedefaultseppunct}\relax
\EndOfBibitem
\bibitem[Yang \latin{et~al.}(2023)Yang, Li, Zhou, Guo, Jia, Liu, Houk, Dubi,
  and Guo]{Yang2023}
Yang,~C.; Li,~Y.; Zhou,~S.; Guo,~Y.; Jia,~C.; Liu,~Z.; Houk,~K.~N.; Dubi,~Y.;
  Guo,~X. Real-time monitoring of reaction stereochemistry through
  single-molecule observations of chirality-induced spin selectivity.
  \emph{Nature Chemistry} \textbf{2023}, \emph{15}, 972--979\relax
\mciteBstWouldAddEndPuncttrue
\mciteSetBstMidEndSepPunct{\mcitedefaultmidpunct}
{\mcitedefaultendpunct}{\mcitedefaultseppunct}\relax
\EndOfBibitem
\bibitem[Mej{\'{\i}}a and Franco(2019)Mej{\'{\i}}a, and Franco]{Mejia2019}
Mej{\'{\i}}a,~L.; Franco,~I. Force{\textendash}conductance spectroscopy of a
  single-molecule reaction. \emph{Chemical Science} \textbf{2019}, \emph{10},
  3249--3256\relax
\mciteBstWouldAddEndPuncttrue
\mciteSetBstMidEndSepPunct{\mcitedefaultmidpunct}
{\mcitedefaultendpunct}{\mcitedefaultseppunct}\relax
\EndOfBibitem
\bibitem[Zang \latin{et~al.}(2019)Zang, Zou, Fu, Ng, Fowler, Yang, Li,
  Steigerwald, Nuckolls, and Venkataraman]{Zang2019a}
Zang,~Y.; Zou,~Q.; Fu,~T.; Ng,~F.; Fowler,~B.; Yang,~J.; Li,~H.;
  Steigerwald,~M.~L.; Nuckolls,~C.; Venkataraman,~L. Directing isomerization
  reactions of cumulenes with electric fields. \emph{Nature Communications}
  \textbf{2019}, \emph{10}, 1--7\relax
\mciteBstWouldAddEndPuncttrue
\mciteSetBstMidEndSepPunct{\mcitedefaultmidpunct}
{\mcitedefaultendpunct}{\mcitedefaultseppunct}\relax
\EndOfBibitem
\bibitem[Ciampi \latin{et~al.}(2018)Ciampi, Darwish, Aitken,
  D{\'{\i}}ez-P{\'{e}}rez, and Coote]{Ciampi2018}
Ciampi,~S.; Darwish,~N.; Aitken,~H.~M.; D{\'{\i}}ez-P{\'{e}}rez,~I.;
  Coote,~M.~L. Harnessing electrostatic catalysis in single molecule,
  electrochemical and chemical systems: a rapidly growing experimental tool
  box. \emph{Chemical Society Reviews} \textbf{2018}, \emph{47},
  5146--5164\relax
\mciteBstWouldAddEndPuncttrue
\mciteSetBstMidEndSepPunct{\mcitedefaultmidpunct}
{\mcitedefaultendpunct}{\mcitedefaultseppunct}\relax
\EndOfBibitem
\bibitem[Huang \latin{et~al.}(2017)Huang, Jevric, Borges, Olsen, Hamill, Zheng,
  Yang, Rudnev, Baghernejad, Broekmann, Petersen, Wandlowski, Mikkelsen,
  Solomon, Nielsen, Hong, Huang, Jevric, Borges, Olsen, Hamill, Zheng, Yang,
  Rudnev, Baghernejad, Broekmann, Petersen, Wandlowski, Mikkelsen, Solomon,
  Nielsen, and Hong]{Huang2017}
Huang,~C. \latin{et~al.}  Single-molecule detection of dihydroazulene
  photo-thermal reaction using break junction technique. \emph{Nature
  Communications} \textbf{2017}, \emph{8}, 1--7\relax
\mciteBstWouldAddEndPuncttrue
\mciteSetBstMidEndSepPunct{\mcitedefaultmidpunct}
{\mcitedefaultendpunct}{\mcitedefaultseppunct}\relax
\EndOfBibitem
\bibitem[Aragon{\`{e}}s \latin{et~al.}(2016)Aragon{\`{e}}s, Haworth, Darwish,
  Ciampi, Bloomfield, Wallace, Diez-Perez, and Coote]{Aragones2016a}
Aragon{\`{e}}s,~A.~C.; Haworth,~N.~L.; Darwish,~N.; Ciampi,~S.;
  Bloomfield,~N.~J.; Wallace,~G.~G.; Diez-Perez,~I.; Coote,~M.~L. Electrostatic
  catalysis of a Diels{\textendash}Alder reaction. \emph{Nature} \textbf{2016},
  \emph{531}, 88--91\relax
\mciteBstWouldAddEndPuncttrue
\mciteSetBstMidEndSepPunct{\mcitedefaultmidpunct}
{\mcitedefaultendpunct}{\mcitedefaultseppunct}\relax
\EndOfBibitem
\bibitem[Besteman \latin{et~al.}(2003)Besteman, Lee, Wiertz, Heering, and
  Dekker]{Besteman2003}
Besteman,~K.; Lee,~J.-O.; Wiertz,~F. G.~M.; Heering,~H.~A.; Dekker,~C.
  Enzyme-Coated Carbon Nanotubes as Single-Molecule Biosensors. \emph{Nano
  Letters} \textbf{2003}, \emph{3}, 727--730\relax
\mciteBstWouldAddEndPuncttrue
\mciteSetBstMidEndSepPunct{\mcitedefaultmidpunct}
{\mcitedefaultendpunct}{\mcitedefaultseppunct}\relax
\EndOfBibitem
\bibitem[Liljeroth \latin{et~al.}(2007)Liljeroth, Repp, and
  Meyer]{Liljeroth2007}
Liljeroth,~P.; Repp,~J.; Meyer,~G. Current-Induced Hydrogen Tautomerization and
  Conductance Switching of Naphthalocyanine Molecules. \emph{Science}
  \textbf{2007}, \emph{317}, 1203--1206\relax
\mciteBstWouldAddEndPuncttrue
\mciteSetBstMidEndSepPunct{\mcitedefaultmidpunct}
{\mcitedefaultendpunct}{\mcitedefaultseppunct}\relax
\EndOfBibitem
\bibitem[O{'}Driscoll\textsuperscript{*}
  \latin{et~al.}(2017)O{'}Driscoll\textsuperscript{*},
  Hamill\textsuperscript{*}, Grace, Nielsen, Almuti, Fu, Hong, Lambert, and
  Jeppesen]{ODriscoll2017}
O{'}Driscoll\textsuperscript{*},~L.~J.; Hamill\textsuperscript{*},~J.~M.;
  Grace,~I.~M.; Nielsen,~B.~W.; Almuti,~E.; Fu,~Y.; Hong,~W.; Lambert,~C.;
  Jeppesen,~J.~O. Electrochemical control of the single molecule conductance of
  a conjugated bis(pyrrolo)tetrathiafulvalene based molecular switch.
  \emph{Chemical Science} \textbf{2017}, \emph{8}, 6123--6130\relax
\mciteBstWouldAddEndPuncttrue
\mciteSetBstMidEndSepPunct{\mcitedefaultmidpunct}
{\mcitedefaultendpunct}{\mcitedefaultseppunct}\relax
\EndOfBibitem
\bibitem[Xu(2014)]{Xu2014}
Xu,~K. Electrolytes and Interphases in Li-Ion Batteries and Beyond.
  \emph{Chemical Reviews} \textbf{2014}, \emph{114}, 11503--11618\relax
\mciteBstWouldAddEndPuncttrue
\mciteSetBstMidEndSepPunct{\mcitedefaultmidpunct}
{\mcitedefaultendpunct}{\mcitedefaultseppunct}\relax
\EndOfBibitem
\bibitem[Li \latin{et~al.}(2012)Li, Mishchenko, and Wandlowski]{Li2012Charge}
Li,~C.; Mishchenko,~A.; Wandlowski,~T. In \emph{Unimolecular and Supramolecular
  Electronics II: Chemistry and Physics Meet at Metal-Molecule Interfaces};
  Metzger,~R.~M., Ed.; Springer Berlin Heidelberg: Berlin, Heidelberg, 2012; pp
  121--188\relax
\mciteBstWouldAddEndPuncttrue
\mciteSetBstMidEndSepPunct{\mcitedefaultmidpunct}
{\mcitedefaultendpunct}{\mcitedefaultseppunct}\relax
\EndOfBibitem
\bibitem[Kamenetska \latin{et~al.}(2009)Kamenetska, Koentopp, Whalley, Park,
  Steigerwald, Nuckolls, Hybertsen, and Venkataraman]{Kamenetska2009}
Kamenetska,~M.; Koentopp,~M.; Whalley,~A.~C.; Park,~Y.~S.; Steigerwald,~M.~L.;
  Nuckolls,~C.; Hybertsen,~M.~S.; Venkataraman,~L. Formation and Evolution of
  Single-Molecule Junctions. \emph{Physical Review Letters} \textbf{2009},
  \emph{102}, 126803\relax
\mciteBstWouldAddEndPuncttrue
\mciteSetBstMidEndSepPunct{\mcitedefaultmidpunct}
{\mcitedefaultendpunct}{\mcitedefaultseppunct}\relax
\EndOfBibitem
\bibitem[Tao(2008)]{Tao2008}
Tao,~F. Nanoscale surface chemistry in self- and directed-assembly of organic
  molecules on solid surfaces and synthesis of nanostructured organic
  architectures. \emph{Pure and Applied Chemistry} \textbf{2008}, \emph{80},
  45--57\relax
\mciteBstWouldAddEndPuncttrue
\mciteSetBstMidEndSepPunct{\mcitedefaultmidpunct}
{\mcitedefaultendpunct}{\mcitedefaultseppunct}\relax
\EndOfBibitem
\bibitem[Hla and Rieder(2003)Hla, and Rieder]{Hla2003}
Hla,~S.-W.; Rieder,~K.-H. {STM} Control of Chemical Reactions: Single-Molecule
  Synthesis. \emph{Annual Review of Physical Chemistry} \textbf{2003},
  \emph{54}, 307--330\relax
\mciteBstWouldAddEndPuncttrue
\mciteSetBstMidEndSepPunct{\mcitedefaultmidpunct}
{\mcitedefaultendpunct}{\mcitedefaultseppunct}\relax
\EndOfBibitem
\bibitem[Huang \latin{et~al.}(2015)Huang, Rudnev, Hong, and
  Wandlowski]{Huang2015break}
Huang,~C.; Rudnev,~A.~V.; Hong,~W.; Wandlowski,~T. Break junction under
  electrochemical gating: testbed for single-molecule electronics.
  \emph{Chemical Society Reviews} \textbf{2015}, \emph{44}, 889--901\relax
\mciteBstWouldAddEndPuncttrue
\mciteSetBstMidEndSepPunct{\mcitedefaultmidpunct}
{\mcitedefaultendpunct}{\mcitedefaultseppunct}\relax
\EndOfBibitem
\bibitem[Mayor and Weber(2004)Mayor, and Weber]{Mayor2004}
Mayor,~M.; Weber,~H.~B. Statistical analysis of single-molecule junctions.
  \emph{Angewandte Chemie, International Edition} \textbf{2004}, \emph{43},
  2882--2884\relax
\mciteBstWouldAddEndPuncttrue
\mciteSetBstMidEndSepPunct{\mcitedefaultmidpunct}
{\mcitedefaultendpunct}{\mcitedefaultseppunct}\relax
\EndOfBibitem
\bibitem[Xu and Tao(2003)Xu, and Tao]{Xu2003Science}
Xu,~B.; Tao,~N.~J. Measurement of Single-Molecule Resistance by Repeated
  Formation of Molecular Junctions. \emph{Science} \textbf{2003}, \emph{301},
  1221--1223\relax
\mciteBstWouldAddEndPuncttrue
\mciteSetBstMidEndSepPunct{\mcitedefaultmidpunct}
{\mcitedefaultendpunct}{\mcitedefaultseppunct}\relax
\EndOfBibitem
\bibitem[Salomon \latin{et~al.}(2003)Salomon, Cahen, Lindsay, Tomfohr,
  Engelkes, and Frisbie]{Salomon2003comparison}
Salomon,~A.; Cahen,~D.; Lindsay,~S.; Tomfohr,~J.; Engelkes,~V.; Frisbie,~C.
  Comparison of Electronic Transport Measurements on Organic Molecules.
  \emph{Advanced Materials} \textbf{2003}, \emph{15}, 1881--1890\relax
\mciteBstWouldAddEndPuncttrue
\mciteSetBstMidEndSepPunct{\mcitedefaultmidpunct}
{\mcitedefaultendpunct}{\mcitedefaultseppunct}\relax
\EndOfBibitem
\bibitem[Solomon \latin{et~al.}(2006)Solomon, Gagliardi, Pecchia, Frauenheim,
  Di~Carlo, Reimers, and Hush]{Solomon2006}
Solomon,~G.~C.; Gagliardi,~A.; Pecchia,~A.; Frauenheim,~T.; Di~Carlo,~A.;
  Reimers,~J.~R.; Hush,~N.~S. Molecular Origins of Conduction Channels Observed
  in Shot-Noise Measurements. \emph{Nano Letters} \textbf{2006}, \emph{6},
  2431--2437\relax
\mciteBstWouldAddEndPuncttrue
\mciteSetBstMidEndSepPunct{\mcitedefaultmidpunct}
{\mcitedefaultendpunct}{\mcitedefaultseppunct}\relax
\EndOfBibitem
\bibitem[Xu \latin{et~al.}(2003)Xu, Xiao, and Tao]{Xu2003JACS}
Xu,~B.; Xiao,~X.; Tao,~N.~J. Measurements of Single-Molecule Electromechanical
  Properties. \emph{Journal of the American Chemical Society} \textbf{2003},
  \emph{125}, 16164--16165\relax
\mciteBstWouldAddEndPuncttrue
\mciteSetBstMidEndSepPunct{\mcitedefaultmidpunct}
{\mcitedefaultendpunct}{\mcitedefaultseppunct}\relax
\EndOfBibitem
\bibitem[Li \latin{et~al.}(2020)Li, Mej{\'{\i}}a, Marrs, Jeong, Hihath, and
  Franco]{Li2020}
Li,~Z.; Mej{\'{\i}}a,~L.; Marrs,~J.; Jeong,~H.; Hihath,~J.; Franco,~I.
  Understanding the Conductance Dispersion of Single-Molecule Junctions.
  \emph{Journal of Physical Chemistry C} \textbf{2020}, \emph{125},
  3406--3414\relax
\mciteBstWouldAddEndPuncttrue
\mciteSetBstMidEndSepPunct{\mcitedefaultmidpunct}
{\mcitedefaultendpunct}{\mcitedefaultseppunct}\relax
\EndOfBibitem
\bibitem[Makk \latin{et~al.}(2012)Makk, Balogh, Csonka, and
  Halbritter]{Makk2012Pulling}
Makk,~P.; Balogh,~Z.; Csonka,~S.; Halbritter,~A. Pulling platinum atomic chains
  by carbon monoxide molecules. \emph{Nanoscale} \textbf{2012}, \emph{4},
  4739--4745\relax
\mciteBstWouldAddEndPuncttrue
\mciteSetBstMidEndSepPunct{\mcitedefaultmidpunct}
{\mcitedefaultendpunct}{\mcitedefaultseppunct}\relax
\EndOfBibitem
\bibitem[Balogh \latin{et~al.}(2015)Balogh, Makk, and
  Halbritter]{balogh2015alternative}
Balogh,~Z.; Makk,~P.; Halbritter,~A. Alternative types of molecule-decorated
  atomic chains in Au--CO--Au single-molecule junctions. \emph{Beilstein
  Journal of Nanotechnology} \textbf{2015}, \emph{6}, 1369\relax
\mciteBstWouldAddEndPuncttrue
\mciteSetBstMidEndSepPunct{\mcitedefaultmidpunct}
{\mcitedefaultendpunct}{\mcitedefaultseppunct}\relax
\EndOfBibitem
\bibitem[Black and Scholes(1973)Black, and Scholes]{Black1973}
Black,~F.; Scholes,~M. The Pricing of Options and Corporate Liabilities.
  \emph{Journal of Political Economy} \textbf{1973}, \emph{81}, 637--654\relax
\mciteBstWouldAddEndPuncttrue
\mciteSetBstMidEndSepPunct{\mcitedefaultmidpunct}
{\mcitedefaultendpunct}{\mcitedefaultseppunct}\relax
\EndOfBibitem
\bibitem[Kirilenko \latin{et~al.}(2017)Kirilenko, Kyle, Samadi, and
  Tuzun]{Kirilenko2017}
Kirilenko,~A.; Kyle,~A.~S.; Samadi,~M.; Tuzun,~T. The flash crash:
  High-frequency trading in an electronic market. \emph{The Journal of Finance}
  \textbf{2017}, \emph{72}, 967--998\relax
\mciteBstWouldAddEndPuncttrue
\mciteSetBstMidEndSepPunct{\mcitedefaultmidpunct}
{\mcitedefaultendpunct}{\mcitedefaultseppunct}\relax
\EndOfBibitem
\bibitem[Not()]{Note-1}
Over 300 securities saw prices decline over \qty{60}{\percent} before
  rebounding – see Ref.~\citenum{Kirilenko2017}\relax
\mciteBstWouldAddEndPuncttrue
\mciteSetBstMidEndSepPunct{\mcitedefaultmidpunct}
{\mcitedefaultendpunct}{\mcitedefaultseppunct}\relax
\EndOfBibitem
\bibitem[Fama(1963)]{Fama1963}
Fama,~E.~F. Mandelbrot and the stable Paretian hypothesis. \emph{The journal of
  business} \textbf{1963}, \emph{36}, 420--429\relax
\mciteBstWouldAddEndPuncttrue
\mciteSetBstMidEndSepPunct{\mcitedefaultmidpunct}
{\mcitedefaultendpunct}{\mcitedefaultseppunct}\relax
\EndOfBibitem
\bibitem[Merton(1976)]{Merton1976}
Merton,~R.~C. Option pricing when underlying stock returns are discontinuous.
  \emph{Journal of Financial Economics} \textbf{1976}, \emph{3}, 125--144\relax
\mciteBstWouldAddEndPuncttrue
\mciteSetBstMidEndSepPunct{\mcitedefaultmidpunct}
{\mcitedefaultendpunct}{\mcitedefaultseppunct}\relax
\EndOfBibitem
\bibitem[Truong \latin{et~al.}(2020)Truong, Oudre, and Vayatis]{Truong2020}
Truong,~C.; Oudre,~L.; Vayatis,~N. Selective review of offline change point
  detection methods. \emph{Signal Processing} \textbf{2020}, \emph{167},
  107299\relax
\mciteBstWouldAddEndPuncttrue
\mciteSetBstMidEndSepPunct{\mcitedefaultmidpunct}
{\mcitedefaultendpunct}{\mcitedefaultseppunct}\relax
\EndOfBibitem
\bibitem[Donayre and Panovska(2021)Donayre, and Panovska]{Donayre2021}
Donayre,~L.; Panovska,~I. Recession-specific recoveries: L's, U's and
  everything in between. \emph{Economics Letters} \textbf{2021}, \emph{209},
  110145\relax
\mciteBstWouldAddEndPuncttrue
\mciteSetBstMidEndSepPunct{\mcitedefaultmidpunct}
{\mcitedefaultendpunct}{\mcitedefaultseppunct}\relax
\EndOfBibitem
\bibitem[Kwiatkowski \latin{et~al.}(1992)Kwiatkowski, Phillips, Schmidt, and
  Shin]{Kwiatkowski1992}
Kwiatkowski,~D.; Phillips,~P.~C.; Schmidt,~P.; Shin,~Y. Testing the null
  hypothesis of stationarity against the alternative of a unit root.
  \emph{Journal of Econometrics} \textbf{1992}, \emph{54}, 159--178\relax
\mciteBstWouldAddEndPuncttrue
\mciteSetBstMidEndSepPunct{\mcitedefaultmidpunct}
{\mcitedefaultendpunct}{\mcitedefaultseppunct}\relax
\EndOfBibitem
\bibitem[Elliott \latin{et~al.}(1992)Elliott, Rothenberg, and
  Stock]{Elliott1992}
Elliott,~G.; Rothenberg,~T.; Stock,~J. \emph{Efficient Tests for an
  Autoregressive Unit Root}; 1992\relax
\mciteBstWouldAddEndPuncttrue
\mciteSetBstMidEndSepPunct{\mcitedefaultmidpunct}
{\mcitedefaultendpunct}{\mcitedefaultseppunct}\relax
\EndOfBibitem
\bibitem[Phillips and Perron(1988)Phillips, and Perron]{PHILLIPS1988}
Phillips,~P. C.~B.; Perron,~P. Testing for a unit root in time series
  regression. \emph{Biometrika} \textbf{1988}, \emph{75}, 335--346\relax
\mciteBstWouldAddEndPuncttrue
\mciteSetBstMidEndSepPunct{\mcitedefaultmidpunct}
{\mcitedefaultendpunct}{\mcitedefaultseppunct}\relax
\EndOfBibitem
\bibitem[Dickey and Fuller(1979)Dickey, and Fuller]{Dickey1979}
Dickey,~D.~A.; Fuller,~W.~A. Distribution of the Estimators for Autoregressive
  Time Series with a Unit Root. \emph{Journal of the American Statistical
  Association} \textbf{1979}, \emph{74}, 427--431\relax
\mciteBstWouldAddEndPuncttrue
\mciteSetBstMidEndSepPunct{\mcitedefaultmidpunct}
{\mcitedefaultendpunct}{\mcitedefaultseppunct}\relax
\EndOfBibitem
\bibitem[Lauritzen \latin{et~al.}(2018)Lauritzen, Magyarkuti, Balogh,
  Halbritter, and Solomon]{Lauritzen2018}
Lauritzen,~K.~P.; Magyarkuti,~A.; Balogh,~Z.; Halbritter,~A.; Solomon,~G.~C.
  Classification of conductance traces with recurrent neural networks.
  \emph{Journal of Chemical Physics} \textbf{2018}, \emph{148}, 084111\relax
\mciteBstWouldAddEndPuncttrue
\mciteSetBstMidEndSepPunct{\mcitedefaultmidpunct}
{\mcitedefaultendpunct}{\mcitedefaultseppunct}\relax
\EndOfBibitem
\bibitem[Magyarkuti \latin{et~al.}(2020)Magyarkuti, Balogh, Balogh,
  Venkataraman, and Halbritter]{Magyarkuti2020}
Magyarkuti,~A.; Balogh,~N.; Balogh,~Z.; Venkataraman,~L.; Halbritter,~A.
  Unsupervised feature recognition in single-molecule break junction data.
  \emph{Nanoscale} \textbf{2020}, \emph{12}, 8355--8363\relax
\mciteBstWouldAddEndPuncttrue
\mciteSetBstMidEndSepPunct{\mcitedefaultmidpunct}
{\mcitedefaultendpunct}{\mcitedefaultseppunct}\relax
\EndOfBibitem
\bibitem[Su \latin{et~al.}(2017)Su, Li, Klausen, Kim, Neupane, Leighton,
  Steigerwald, Venkataraman, and Nuckolls]{Su2017}
Su,~T.~A.; Li,~H.; Klausen,~R.~S.; Kim,~N.~T.; Neupane,~M.; Leighton,~J.~L.;
  Steigerwald,~M.~L.; Venkataraman,~L.; Nuckolls,~C. Silane and Germane
  Molecular Electronics. \emph{Accounts of Chemical Research} \textbf{2017},
  \emph{50}, 1088--1095\relax
\mciteBstWouldAddEndPuncttrue
\mciteSetBstMidEndSepPunct{\mcitedefaultmidpunct}
{\mcitedefaultendpunct}{\mcitedefaultseppunct}\relax
\EndOfBibitem
\bibitem[van~den Burg and Williams(2020)van~den Burg, and Williams]{Burg2020}
van~den Burg,~G. J.~J.; Williams,~C. K.~I. An Evaluation of Change Point
  Detection Algorithms. \textbf{2020}, \relax
\mciteBstWouldAddEndPunctfalse
\mciteSetBstMidEndSepPunct{\mcitedefaultmidpunct}
{}{\mcitedefaultseppunct}\relax
\EndOfBibitem
\bibitem[Truong \latin{et~al.}(2018)Truong, Oudre, and Vayatis]{Truong2018}
Truong,~C.; Oudre,~L.; Vayatis,~N. ruptures: change point detection in Python.
  \textbf{2018}, \relax
\mciteBstWouldAddEndPunctfalse
\mciteSetBstMidEndSepPunct{\mcitedefaultmidpunct}
{}{\mcitedefaultseppunct}\relax
\EndOfBibitem
\bibitem[Soler \latin{et~al.}(2002)Soler, Artacho, Gale, Garc{\'\i}a, Junquera,
  Ordej{\'o}n, and S{\'a}nchez-Portal]{Soler2002}
Soler,~J.~M.; Artacho,~E.; Gale,~J.~D.; Garc{\'\i}a,~A.; Junquera,~J.;
  Ordej{\'o}n,~P.; S{\'a}nchez-Portal,~D. The SIESTA method for {\textit{ab
  initio}} order-{\textit{N}} materials simulation. \emph{Journal of Physics:
  Condensed Matter} \textbf{2002}, \emph{14}, 2745\relax
\mciteBstWouldAddEndPuncttrue
\mciteSetBstMidEndSepPunct{\mcitedefaultmidpunct}
{\mcitedefaultendpunct}{\mcitedefaultseppunct}\relax
\EndOfBibitem
\bibitem[Brandbyge \latin{et~al.}(2002)Brandbyge, Mozos, Ordej{\'{o}}n, Taylor,
  and Stokbro]{Brandbyge2002}
Brandbyge,~M.; Mozos,~J.-L.; Ordej{\'{o}}n,~P.; Taylor,~J.; Stokbro,~K.
  Density-functional method for nonequilibrium electron transport.
  \emph{Physical Review B} \textbf{2002}, \emph{65}, 165401\relax
\mciteBstWouldAddEndPuncttrue
\mciteSetBstMidEndSepPunct{\mcitedefaultmidpunct}
{\mcitedefaultendpunct}{\mcitedefaultseppunct}\relax
\EndOfBibitem
\bibitem[Papior \latin{et~al.}(2017)Papior, Lorente, Frederiksen,
  Garc{\'{\i}}a, and Brandbyge]{Papior2017}
Papior,~N.; Lorente,~N.; Frederiksen,~T.; Garc{\'{\i}}a,~A.; Brandbyge,~M.
  Improvements on non-equilibrium and transport Green function techniques: The
  next-generation transiesta. \emph{Computer Physics Communications}
  \textbf{2017}, \emph{212}, 8--24\relax
\mciteBstWouldAddEndPuncttrue
\mciteSetBstMidEndSepPunct{\mcitedefaultmidpunct}
{\mcitedefaultendpunct}{\mcitedefaultseppunct}\relax
\EndOfBibitem
\bibitem[Perdew \latin{et~al.}(1996)Perdew, Burke, and Ernzerhof]{Perdew1996}
Perdew,~J.~P.; Burke,~K.; Ernzerhof,~M. Generalized Gradient Approximation Made
  Simple. \emph{Physical Review Letters} \textbf{1996}, \emph{77},
  3865--3868\relax
\mciteBstWouldAddEndPuncttrue
\mciteSetBstMidEndSepPunct{\mcitedefaultmidpunct}
{\mcitedefaultendpunct}{\mcitedefaultseppunct}\relax
\EndOfBibitem
\bibitem[Kaliginedi \latin{et~al.}(2014)Kaliginedi, V.~Rudnev,
  Moreno-Garc{\'i}a, Baghernejad, Huang, Hong, and
  Wandlowski]{Kaliginedi2014promising}
Kaliginedi,~V.; V.~Rudnev,~A.; Moreno-Garc{\'i}a,~P.; Baghernejad,~M.;
  Huang,~C.; Hong,~W.; Wandlowski,~T. Promising anchoring groups for
  single-molecule conductance measurements. \emph{Physical Chemistry Chemical
  Physics} \textbf{2014}, \emph{16}, 23529--23539\relax
\mciteBstWouldAddEndPuncttrue
\mciteSetBstMidEndSepPunct{\mcitedefaultmidpunct}
{\mcitedefaultendpunct}{\mcitedefaultseppunct}\relax
\EndOfBibitem
\bibitem[Borges \latin{et~al.}(2016)Borges, Fung, Ng, Venkataraman, and
  Solomon]{Borges2016}
Borges,~A.; Fung,~E.-D.; Ng,~F.; Venkataraman,~L.; Solomon,~G.~C. Probing the
  Conductance of the $\sigma$-System of Bipyridine Using Destructive
  Interference. \emph{Journal of Physical Chemistry Letters} \textbf{2016},
  \emph{7}, 4825--4829\relax
\mciteBstWouldAddEndPuncttrue
\mciteSetBstMidEndSepPunct{\mcitedefaultmidpunct}
{\mcitedefaultendpunct}{\mcitedefaultseppunct}\relax
\EndOfBibitem
\bibitem[Hamill \latin{et~al.}(2014)Hamill, Wang, and Xu]{Hamill2014Nanoscale}
Hamill,~J.~M.; Wang,~K.; Xu,~B. Force and conductance molecular break junctions
  with time series crosscorrelation. \emph{Nanoscale} \textbf{2014}, \emph{6},
  5657--5661\relax
\mciteBstWouldAddEndPuncttrue
\mciteSetBstMidEndSepPunct{\mcitedefaultmidpunct}
{\mcitedefaultendpunct}{\mcitedefaultseppunct}\relax
\EndOfBibitem
\bibitem[P\'{e}rez-Jim\'{e}nez \latin{et~al.}(2005)P\'{e}rez-Jim\'{e}nez,
  Sancho-Garc\'{\i}a, and P{\'e}rez-Jord{\'a}]{Perez-Jimenez2005JCP}
P\'{e}rez-Jim\'{e}nez,~A.~J.; Sancho-Garc\'{\i}a,~J.~C.;
  P{\'e}rez-Jord{\'a},~J.~M. Torsional potential of 4,4'-bipyridine: Ab initio
  analysis of dispersion and vibrational effects. \emph{Journal of Chemical
  Physics} \textbf{2005}, \emph{123}, 134309\relax
\mciteBstWouldAddEndPuncttrue
\mciteSetBstMidEndSepPunct{\mcitedefaultmidpunct}
{\mcitedefaultendpunct}{\mcitedefaultseppunct}\relax
\EndOfBibitem
\bibitem[Makk \latin{et~al.}(2012)Makk, Tomaszewski, Martinek, Balogh, Csonka,
  Wawrzyniak, Frei, Venkataraman, and Halbritter]{Makk2012}
Makk,~P.; Tomaszewski,~D.; Martinek,~J.; Balogh,~Z.; Csonka,~S.;
  Wawrzyniak,~M.; Frei,~M.; Venkataraman,~L.; Halbritter,~A. Correlation
  Analysis of Atomic and Single-Molecule Junction Conductance. \emph{ACS Nano}
  \textbf{2012}, \emph{6}, 3411--3423\relax
\mciteBstWouldAddEndPuncttrue
\mciteSetBstMidEndSepPunct{\mcitedefaultmidpunct}
{\mcitedefaultendpunct}{\mcitedefaultseppunct}\relax
\EndOfBibitem
\bibitem[Quek \latin{et~al.}(2009)Quek, Kamenetska, Steigerwald, Choi, Louie,
  Hybertsen, Neaton, and Venkataraman]{Quek2009NN}
Quek,~S.~Y.; Kamenetska,~M.; Steigerwald,~M.~L.; Choi,~H.~J.; Louie,~S.~G.;
  Hybertsen,~M.~S.; Neaton,~J.~B.; Venkataraman,~L. Mechanically controlled
  binary conductance switching of a single-molecule junction. \emph{Nature
  Nanotechnology} \textbf{2009}, \emph{4}, 230--234\relax
\mciteBstWouldAddEndPuncttrue
\mciteSetBstMidEndSepPunct{\mcitedefaultmidpunct}
{\mcitedefaultendpunct}{\mcitedefaultseppunct}\relax
\EndOfBibitem
\bibitem[Savitzky and Golay(1964)Savitzky, and Golay]{Savitzky1964}
Savitzky,~A.; Golay,~M. J.~E. Smoothing and Differentiation of Data by
  Simplified Least Squares Procedures. \emph{Analytical Chemistry}
  \textbf{1964}, \emph{36}, 1627--1639\relax
\mciteBstWouldAddEndPuncttrue
\mciteSetBstMidEndSepPunct{\mcitedefaultmidpunct}
{\mcitedefaultendpunct}{\mcitedefaultseppunct}\relax
\EndOfBibitem
\bibitem[Virtanen \latin{et~al.}(2020)Virtanen, Gommers, Oliphant, Haberland,
  Reddy, Cournapeau, Burovski, Peterson, Weckesser, Bright, {van der Walt},
  Brett, Wilson, Millman, Mayorov, Nelson, Jones, Kern, Larson, Carey, Polat,
  Feng, Moore, {VanderPlas}, Laxalde, Perktold, Cimrman, Henriksen, Quintero,
  Harris, Archibald, Ribeiro, Pedregosa, {van Mulbregt}, and {SciPy 1.0
  Contributors}]{2020SciPy_NMeth}
Virtanen,~P. \latin{et~al.}  {{SciPy} 1.0: Fundamental Algorithms for
  Scientific Computing in Python}. \emph{Nature Methods} \textbf{2020},
  \emph{17}, 261--272\relax
\mciteBstWouldAddEndPuncttrue
\mciteSetBstMidEndSepPunct{\mcitedefaultmidpunct}
{\mcitedefaultendpunct}{\mcitedefaultseppunct}\relax
\EndOfBibitem
\bibitem[Fujihira \latin{et~al.}(2006)Fujihira, Suzuki, Fujii, and
  Nishikawa]{Fujihira2006}
Fujihira,~M.; Suzuki,~M.; Fujii,~S.; Nishikawa,~A. Currents through single
  molecular junction of Au/hexanedithiolate/Au measured by repeated formation
  of break junction in {STM} under {UHV}: Effects of conformational change in
  an alkylene chain from gauche to trans and binding sites of thiolates on
  gold. \emph{Physical Chemistry Chemical Physics} \textbf{2006}, \emph{8},
  3876\relax
\mciteBstWouldAddEndPuncttrue
\mciteSetBstMidEndSepPunct{\mcitedefaultmidpunct}
{\mcitedefaultendpunct}{\mcitedefaultseppunct}\relax
\EndOfBibitem
\end{mcitethebibliography}


\providecommand{\latin}[1]{#1}
\makeatletter
\providecommand{\doi}
  {\begingroup\let\do\@makeother\dospecials
  \catcode`\{=1 \catcode`\}=2 \doi@aux}
\providecommand{\doi@aux}[1]{\endgroup\texttt{#1}}
\makeatother
\providecommand*\mcitethebibliography{\thebibliography}
\csname @ifundefined\endcsname{endmcitethebibliography}
  {\let\endmcitethebibliography\endthebibliography}{}
\begin{mcitethebibliography}{5}
\providecommand*\natexlab[1]{#1}
\providecommand*\mciteSetBstSublistMode[1]{}
\providecommand*\mciteSetBstMaxWidthForm[2]{}
\providecommand*\mciteBstWouldAddEndPuncttrue
  {\def\EndOfBibitem{\unskip.}}
\providecommand*\mciteBstWouldAddEndPunctfalse
  {\let\EndOfBibitem\relax}
\providecommand*\mciteSetBstMidEndSepPunct[3]{}
\providecommand*\mciteSetBstSublistLabelBeginEnd[3]{}
\providecommand*\EndOfBibitem{}
\mciteSetBstSublistMode{f}
\mciteSetBstMaxWidthForm{subitem}{(\alph{mcitesubitemcount})}
\mciteSetBstSublistLabelBeginEnd
  {\mcitemaxwidthsubitemform\space}
  {\relax}
  {\relax}

\bibitem[Soler \latin{et~al.}(2002)Soler, Artacho, Gale, Garc{\'\i}a, Junquera,
  Ordej{\'o}n, and S{\'a}nchez-Portal]{Soler2002}
Soler,~J.~M.; Artacho,~E.; Gale,~J.~D.; Garc{\'\i}a,~A.; Junquera,~J.;
  Ordej{\'o}n,~P.; S{\'a}nchez-Portal,~D. The SIESTA method for {\textit{ab
  initio}} order-{\textit{N}} materials simulation. \emph{Journal of Physics:
  Condensed Matter} \textbf{2002}, \emph{14}, 2745\relax
\mciteBstWouldAddEndPuncttrue
\mciteSetBstMidEndSepPunct{\mcitedefaultmidpunct}
{\mcitedefaultendpunct}{\mcitedefaultseppunct}\relax
\EndOfBibitem
\bibitem[Perdew \latin{et~al.}(1996)Perdew, Burke, and Ernzerhof]{Perdew1996}
Perdew,~J.~P.; Burke,~K.; Ernzerhof,~M. Generalized Gradient Approximation Made
  Simple. \emph{Physical Review Letters} \textbf{1996}, \emph{77},
  3865--3868\relax
\mciteBstWouldAddEndPuncttrue
\mciteSetBstMidEndSepPunct{\mcitedefaultmidpunct}
{\mcitedefaultendpunct}{\mcitedefaultseppunct}\relax
\EndOfBibitem
\bibitem[Brandbyge \latin{et~al.}(2002)Brandbyge, Mozos, Ordej{\'{o}}n, Taylor,
  and Stokbro]{Brandbyge2002}
Brandbyge,~M.; Mozos,~J.-L.; Ordej{\'{o}}n,~P.; Taylor,~J.; Stokbro,~K.
  Density-functional method for nonequilibrium electron transport.
  \emph{Physical Review B} \textbf{2002}, \emph{65}, 165401\relax
\mciteBstWouldAddEndPuncttrue
\mciteSetBstMidEndSepPunct{\mcitedefaultmidpunct}
{\mcitedefaultendpunct}{\mcitedefaultseppunct}\relax
\EndOfBibitem
\bibitem[Papior \latin{et~al.}(2017)Papior, Lorente, Frederiksen,
  Garc{\'{\i}}a, and Brandbyge]{Papior2017}
Papior,~N.; Lorente,~N.; Frederiksen,~T.; Garc{\'{\i}}a,~A.; Brandbyge,~M.
  Improvements on non-equilibrium and transport Green function techniques: The
  next-generation transiesta. \emph{Computer Physics Communications}
  \textbf{2017}, \emph{212}, 8--24\relax
\mciteBstWouldAddEndPuncttrue
\mciteSetBstMidEndSepPunct{\mcitedefaultmidpunct}
{\mcitedefaultendpunct}{\mcitedefaultseppunct}\relax
\EndOfBibitem
\end{mcitethebibliography}

\end{document}


\section{Theoretical simulations to illustrate conductance-displacement trends in two target molecules}

We performed simulations of the molecular transport properties in this Letter using SIESTA.\cite{Soler2002}
SIESTA adopts a localized basis set approach within the framework of \ac{dft}.
The \ac{pbe} exchange-correlation functional in the \ac{gga}\cite{Perdew1996} was employed.
All atoms were modeled using a \ac{dzp} basis set, with the exception of Au atoms which were treated using a \ac{dz} basis  set.
The real-space grid for numerical integration was set to have a cutoff of \qty{300}{\rydberg}.
The Brillouin zone was sampled using a Monkhorst-Pack k-point grid of $6\times6\times1$.
The \ac{scf} convergence criteria were set to a density matrix tolerance of $1\times10^{-4}$.
A linear mixing scheme with a weight of \num{0.10} and a history of \num{12} iterations was implemented for \ac{scf} convergence.
Geometry constraints were applied to all Au atoms during geometry optimization.
The geometry was optimized using a conjugate gradient method with a maximum force tolerance of \qty{0.05}{\eV \per \angstrom}.

Utilizing the non-equilibrium Green’s function framework, zero-bias conductance was calculated with TranSIESTA.\cite{Brandbyge2002, Papior2017}
TranSIESTA employs periodic boundary conditions in the xy-plane and designates the z-direction for transport.
The calculation was set up using a $4\times4$ slab of Au(111) atoms to act as electrodes.
The electronic structure of the electrodes were calculated using a Monkhorst-Pack k-point grid of $24\times24\times100$.
The studied molecules were positioned between the flat surfaces of these Au electrodes.
The electronic transmission was subsequently computed using the Landauer formalism.
A denser $16\times16$ k-mesh was employed in a post-processing step for this purpose.

Optimized structures of molecules and input files are available at \\https://doi.org/10.19061/iochem-bd-9-7.

\section{Sensitivity testing of four parameters used in the Chow+ test}\label{sssec:sensitivity}
\begin{figure}[]%
	\includegraphics[width=\figurewidth,keepaspectratio,]{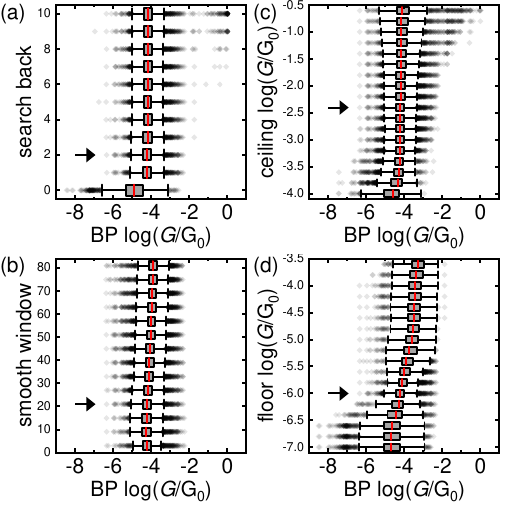}
	\caption{Sensitivity tests of the four parameters required to determine \acfp{bp} using Chow+ summarized in box plots applied to \ac{bpy} data at \qty{4}{\K}. Results of the search window (a) backward search size, (b) smoothing parameter, (c) ceiling, and (d) floor. Boxes are \numrange{25}{75} percentiles, whiskers are \num{1.5} standard deviations, and vertical line is the distribution median; outliers plotted individually. Black arrows show final choice for parameter value.}%
	\label{fig:sensitivity}%
\end{figure}

\begin{figure}[]%
	\includegraphics[width=\figurewidth,keepaspectratio,]{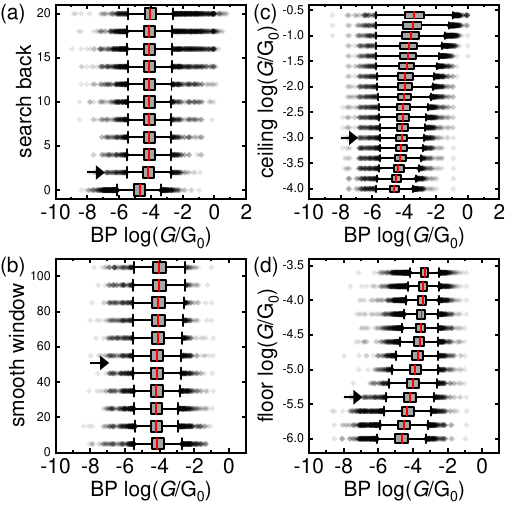}
	\caption{Sensitivity tests of the four parameters required to determine \acfp{bp} using Chow+ summarized in box plots applied to \ac{bpy} data at \acl{rt}. Results of the search window (a) backward search size, (b) smoothing parameter, (c) ceiling, and (d) floor. Boxes are \numrange{25}{75} percentiles, whiskers are \num{1.5} standard deviations, and vertical line is the distribution median; outliers plotted individually. Black arrows show final choice for parameter value.}%
	\label{fig:sensitivityRT}%
\end{figure}

\begin{figure}[]%
	\includegraphics[width=\figurewidth,keepaspectratio,]{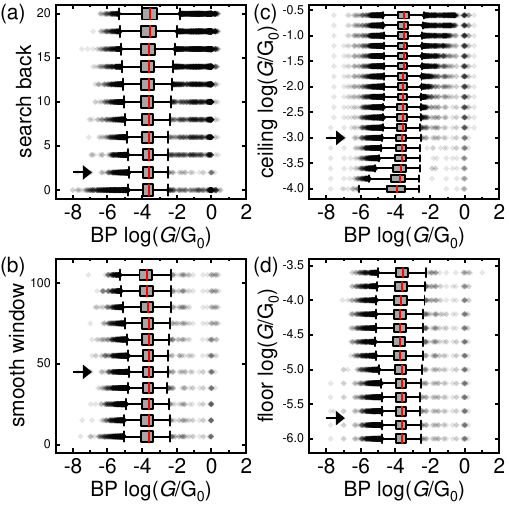}
	\caption{Sensitivity tests of the four parameters required to determine \acfp{bp} using Chow+ summarized in box plots applied to \ac{si4} data. Results of the search window (a) backward search size, (b) smoothing parameter, (c) ceiling, and (d) floor. Boxes are \numrange{25}{75} percentiles, whiskers are \num{1.5} standard deviations, and vertical line is the distribution median; outliers plotted individually. Black arrows show final choice for parameter value.}%
	\label{fig:sensitivitySi}%
\end{figure}

It was important to understand how the \acl{bp} from the Chow test was dependent on the choice of parameters.
The four parameters necessary to find the \acl{bp} presented in the main text were
\begin{enumerate*}[(i)]
	\item number of backward points to search,
	\item smoothing parameter,
	\item search window ceiling, and
	\item search window floor.
\end{enumerate*}
To test the sensitivity of the results to each parameter, three of the parameters were held constant, and the fourth was swept through a linear range of possible values.
In Fig.~\ref{fig:sensitivity}(a), the smoothing and window ceiling and floor were held constant at \num{21}, \qty{}{10^{-2.4}~\G}, and \qty{}{10^{-6.0}~\G}, respectively, and the search back length, in data points, was varied from \numrange{0}{10}.
The resulting distribution of \aclp{bp} for each search back length were then summarized using box plots, where the boxes span the \numrange{25}{75} percentiles, the vertical line marks the median, and the whiskers span the \num{1.5} standard deviation window; outliers are marked individually.
In all the datasets studied herein it was determined that \num{2} data points was sufficient in each case, as shown in Figs.~\ref{fig:sensitivity}-\ref{fig:sensitivitySi}(a).

The smoothing window size was heavily dependent on the sampling rates for the individual datasets, because when the sampling rate was high, as was the case with the \ac{si4}, fluctuations in the signal due simply to the high sampling rate were large enough to impact the Chow test.
However, as can be seen in Figs.~\ref{fig:sensitivity}-\ref{fig:sensitivitySi}(b), the choice had little impact on $\overline{G}_{\text{mol}}$, but only on the variance in the distribution of \aclp{bp}. The values chosen for the size of the smoothing window were \num{21}, \num{51}, and \num{45} for \ac{bpy} at \qty{4}{\kelvin} and \qty{300}{\kelvin}, and \ac{si4}, respectively.

The windowing ceiling and floor we explored qualitatively in Fig. 2 of the main text.
For the \ac{bpy} data at \qty{4}{\K}, in Fig.~\ref{fig:sensitivity}(c), the \acl{bp} window ceiling was swept from \qty{}{10^{-0.4}~\G} to \qty{}{10^{-4.0}~\G} in steps of \num{0.2} on the $\log G/\text{G}_0$ scale.
Because no strong dependency was noticed, the value of \qty{}{10^{-2.4}~\G} was chosen as the ceiling threshold because it ignored superfluous data in the snap-back region.

In Fig.~\ref{fig:sensitivity}(b) the \acl{bp} search window floor was swept from \qty{}{10^{-7.0}~\G} to \qty{}{10^{-3.6}~\G} in steps of \num{0.2} on the $\log G/\text{G}_0$ scale.
In this case, there appeared to be a sensitivity to the choice of window floor.
If the floor was chosen too low, in the noise region below \qty{}{10^{-6.3}~\G}, the \aclp{bp} were frequently chosen within the noise region, where we do not expect to observe a \acl{bp}.
Conversely, if the floor was chosen too high, \qty{}{>10^{-5.0}~\G}, the \aclp{bp} were predominantly chosen to be above the molecular plateau.
The choice of \qty{}{10^{-6.0}~\G} was close to, but not within, the noise floor, and appeared to minimize the variance of the \aclp{bp}.

For \ac{bpy} at \acl{rt}, the Chow+ test was much more sensitive to choice of window ceiling and floor.
A compromise was chosen with a ceiling at \qty{}{10^{-3.0}~\G} and a floor at \qty{}{10^{-5.4}~\G}, because, although the narrower window was more biased, it ensured that reasonable \aclp{bp} were found with Chow+.
The smoothing window had little effect on the results, but better results were achieved with a smoothing window at \num{51} data points.

There was little dependence of the \ac{si4} \aclp{bp} on window floor, so this was set as close to the noise floor as possible, at \qty{}{10^{-5.7}~\G}.
There was some dependence on window ceiling, however. This value was set to just above the plateau region at \qty{}{10^{-3.0}~\G}.

\bibliography{library-break_point_v1-10-acs.bib}